\documentclass{elsart}

\usepackage[american]{babel}
\usepackage[isolatin]{inputenc}         
\usepackage{subfigure}
\usepackage{epsfig}
\usepackage{amssymb,amsmath,amsfonts}
\usepackage[]{graphicx}
\graphicspath{{Figures/}}

\usepackage{natbib}
\newcommand{\xx}{{\bf x}}
\newcommand{\yy}{{\bf y}}
\newcommand{\dxt}{\frac{d{x}}{d{t}}}
\newcommand{\dxxtt}{\frac{d{\xx}}{d{\tau}}}
\newcommand{\dyt}{\frac{d{y}}{d{t}}}
\newcommand{\dyytt}{\frac{d{\yy}}{d{\tau}}}
\newcommand{\ts}{{\tilde s}}
\newcommand{\tu}{{\tilde u}}
\newcommand{\be}{\begin{equation}}
\newcommand{\ee}{\end{equation}}

\begin{document}

\begin{frontmatter}
  \title{Towards an understanding of lineage specification in
    hematopoietic stem cells:\\ A mathematical model for the
    interaction of transcription factors GATA-1 and PU.1 }

\author{Ingo Roeder\corauthref{cor1}}
\ead{ingo.roeder@imise.uni-leipzig.de} 
\author{Ingmar Glauche} 
\ead{ingmar.glauche@imise.uni-leipzig.de}
\address{Institute for Medical Informatics, Statistics and
  Epidemiology, University of Leipzig, Haertelstr. 16/18, D-04107
  Leipzig, Germany}

\corauth[cor1]{Corresponding author. Tel.: +49 (0)341 97 16 111; fax:
  +49 (0)341 97 16 109}

\begin{abstract}
  In addition to their self-renewal capabilities, hematopoietic stem
  cells guarantee the continuous supply of fully differentiated,
  functional cells of various types in the peripheral blood. The
  process which controls differentiation into the different lineages
  of the hematopoietic system (erythroid, myeloid, lymphoid) is
  referred to as lineage specification. It requires a potentially
  multi-step decision sequence which determines the fate of the cells
  and their successors. It is generally accepted that lineage
  specification is regulated by a complex system of interacting
  transcription factors.  However, the underlying principles
  controlling this regulation are currently unknown.

  Here, we propose a simple quantitative model describing the
  interaction of two transcription factors. This model is motivated by
  experimental observations on the transcription factors GATA-1 and
  PU.1, both known to act as key regulators and potential antagonists
  in the erythroid vs. myeloid differentiation processes of
  hematopoietic progenitor cells. We demonstrate the ability of the
  model to account for the observed switching behavior of a transition
  from a state of low expression of both factors (undifferentiated
  state) to the dominance of one factor (differentiated state).
  Depending on the parameter choice, the model predicts two different
  possibilities to explain the experimentally suggested, stem cell
  characterizing \emph{priming} state of low level co-expression.
  Whereas increasing transcription rates are sufficient to induce
  differentiation in one scenario, an additional system perturbation
  (by stochastic fluctuations or directed impulses) of transcription
  factor levels is required in the other case.
  \end{abstract}


\end{frontmatter}

\newpage
\section{Introduction}\label{intro}

The hematopoietic system consists of a variety of functionally
different cell types, including mature cells such as erythrocytes,
granulocytes, platelets, or lymphocytes, as well as several different
precursor cells (i.e., premature cell stages) and hematopoietic stem
cells (HSC) \citep{lord-b-1997-401-a, orkin-s-2000-57-a}. Most mature
cell types have limited life spans ranging from a few hours to several
months, which implies the existence of a source capable of
replenishing these differentiated cells throughout the life span of an
individual. This supply is realized by the population of HSC, which is
maintained and even regenerated after injury or depletion throughout
the whole life of the organism.  This self-renewal property is a major
characteristic defining HSC \citep{loeffler-m-2002-8-a,
  lord-b-1997-401-a,potten-c-1990-1001-a}. A second major
characteristic of HSC is their ability to contribute to the production
of cells of all hematopoietic lineages, thus ensuring the supply of
functionally differentiated cells meeting the needs of the organism.
The process controlling the development of undifferentiated stem or
progenitor cells into one specific functional direction (i.e., one
specific hematopoietic lineage) is called \emph{lineage
  specification}.  It is generally accepted that the process of
lineage specification is governed by the interplay of many different
transcription factors \citep{akashi-k-2005-125-a,cantor-a-2002-3368-a,
  cross-m-1994-3013-a, orkin-s-1995-870-a,
  orkin-s-2000-57-a,tenen-d-2003-89-a}. Experimental results suggest
that a number of relevant transcription factors are expressed
simultaneously in HSC, although at a low level
\citep{akashi-k-2003-383-a,hu-m-1997-774-a}.  Some authors refer to
this state of a low level co-expression as \emph{priming} behavior
\citep{akashi-k-2005-125-a,cross-m-1997-609-a, enver-t-1998-9-a}.
During differentiation the balanced co-expression of these potentially
antagonistic transcription factors is assumed to be broken at some
point (or even multiple points). Thereafter, the system is supposed to
be characterized by an up-regulated level of some transcription
factors, specific for a particular lineage, while other transcription
factors are down-regulated. These observations suggest a transcription
factor network, capable of switch-like behavior by changing from
unspecific co-expression to different states of specific expression.
However, the general underlying principles of the regulatory
mechanisms are currently unknown. Particularly, it is unclear whether
the assumption of a dynamically balanced low level co-expression state
is justified or whether \emph{priming} should rather be interpreted as
the result of an inactive transcription factor network overlaid by
stochastic fluctuations of transcription factor expression.

In this paper we propose a simple mathematical model describing
different interaction scenarios of two transcription factors.
Biologically, this simple two component network model is motivated by
experimental observations on the transcription factors GATA-1 and
PU.1, known to be involved in the process of lineage specification of
HSC \citep{du-j-2002-43481-a,oikawa-t-1999-599-a,
  rekhtman-n-1999-1398-a,rosmarin-a-2005-131-a,
  tenen-d-2003-89-a,voso-m-1994-7932-a}.  The zinc finger factor
GATA-1 is reported to be required for the differentiation and
maturation of erythroid/megakaryocytic cells, while the Ets-family
transcription factor PU.1 supports the development of myeloid and
lymphoid cells \citep[reviewed
by][]{cantor-a-2002-3368-a,tenen-d-2003-89-a}. For both, GATA-1 and
PU.1, it has been demonstrated that they are able to stimulate their
own transcription through an auto-catalytic process
\citep{chen-h-1995-1549-a, nishimura-s-2000-713-a,okuno-y-2005-2832-a,
  tsai-s-1991-919-a}.  Additionally, there are physical interactions
between GATA-1 and PU.1 which induce a mutual inhibition and,
therefore, favor one lineage choice at the expense of the other
(erythroid/megakaryocyte vs.  myeloid) \citep{du-j-2002-43481-a,
  nerlov-c-2000-2543-a,rekhtman-n-1999-1398-a, rekhtman-n-2003-7460-a,
  voso-m-1994-7932-a, yamada-t-1998-186-a, zhang-p-1999-8705-a,
  zhang-p-2000-2641-a}.  In particular, two different mechanisms for
the mutual inhibition of these two transcription factors have been
suggested by experimental observations: On one hand, GATA-1 binds to
the $\beta 3/\beta 4$ region of PU.1 (complex 1) and displaces the
PU.1 co-activator c-Jun from its binding site, thereby, inhibiting the
transcription initiation of PU.1 \citep{zhang-p-1999-8705-a}. On the
other hand, the inhibition of GATA-1 transcription is mediated by the
binding of the N-terminal region of PU.1 to the C-finger region of
GATA-1 (complex 2), thus blocking the binding of GATA-1 to its
promoter \citep{zhang-p-2000-2641-a}. That means, although both
inhibition mechanisms are interfered through the formation of
PU.1/GATA-1 heterodimers, the two complexes are structurally
different. Whereas complex 1 (inhibition of PU.1 transcription by
GATA-1) is known to bind to DNA, thus occupying a PU.1 promoter site,
DNA-binding of complex 2 (inhibition of GATA-1 transcription by PU.1)
has not been reported so far.

The mechanisms of antagonistic interdependence together with positive
auto-catalytic regulation provide a framework for the theoretical
investigation of different scenarios of transcription factor
interaction and their implications for the explanation of lineage
specification control. Applying a mathematical model, which formalizes
the described interactions, it is now possible to analyze different
combinations of transcription factor activation and inhibition on a
qualitative and quantitative level. The proposed model relies on
principles suggested for the description of general genetic switches
\citep[e.g.][]{becskei-a-2001-2528-a, cinquin-o-2002-216-a,
  cinquin-o-2005-233-a,gardner-t-2000-339-a}.

In this paper it is our objective to examine the following questions
within the framework of this model structure: 
\begin{itemize}
\item Are the experimentally described interactions of the two
  transcription factors sufficient to generate a switching behavior
  between a stable co-expression of two factors and the dominance of
  one of these factors?
\item What are the conditions inducing such a qualitative change in
  the system behavior?
\item Is there evidence for a functional role of the (experimentally
  suggested) \emph{priming} status?
\end{itemize}

To answer these question the following strategy is applied.  Firstly,
the model equations are derived on the basis of the described
biological mechanisms of transcription factor interaction for GATA-1
and PU.1 (Section \ref{model}). Secondly, this model is analyzed with
respect to the existence of steady state solutions and their
dependence on the model parameters. According to our objective, to
understand the mechanisms leading to switches between different stable
system states, we focus our analysis particularly on the determination
of bifurcation conditions, considering different scenarios of
transcription factor interaction (Section \ref{res}). Finally, the
obtained results are discussed in relation to the ongoing debate about
lineage specification control in the hematopoietic system,
specifically with respect to potential explanations of the
experimentally suggested low level co-expression of transcription
factors (\emph{priming}) in undifferentiated progenitors and stem
cells (Section \ref{diss}).

\section{Model description}\label{model}

Although our analysis is motivated by experimental observations of
specific transcription factor interactions (GATA-1 and PU.1), the
model may also be applied in the general context of two
interacting transcription factors. In the following, the two
transcription factors are denoted by $X$ and $Y$.

\subsection{General assumptions}\label{model:general}

\begin{figure}
\begin{center}
\subfigure[]{\includegraphics[clip,width=5.5cm]{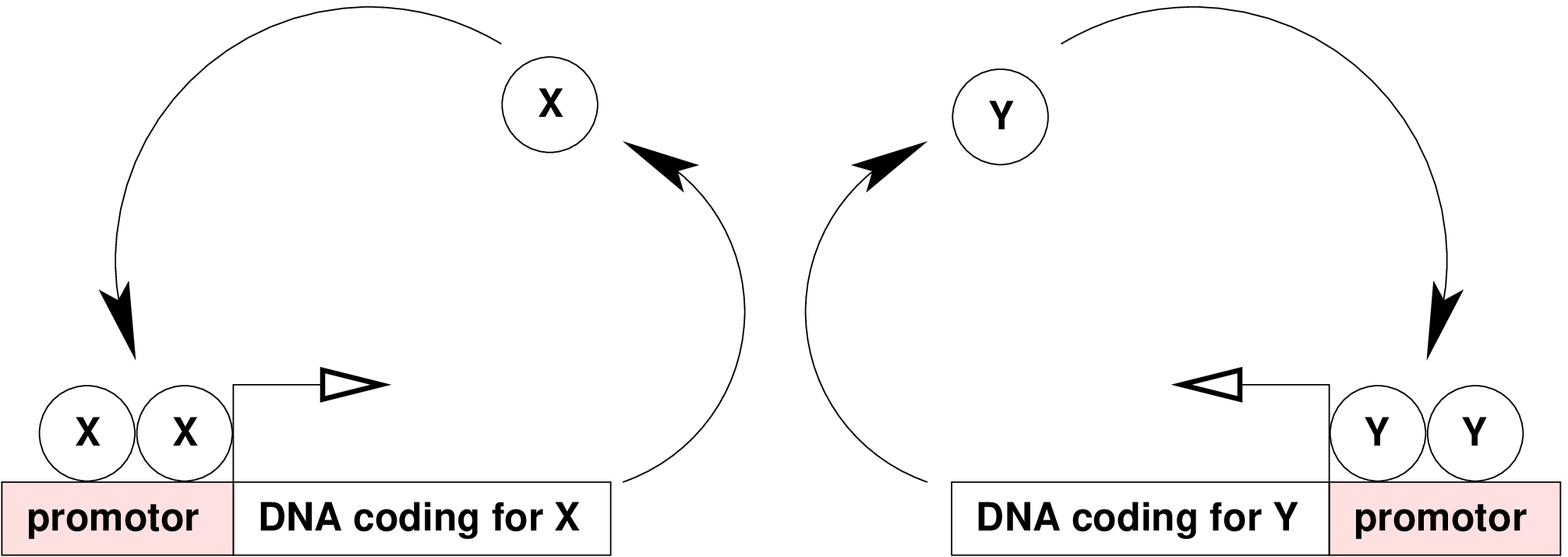}}
\hspace{1cm}
\subfigure[]{\includegraphics[clip,width=5.5cm]{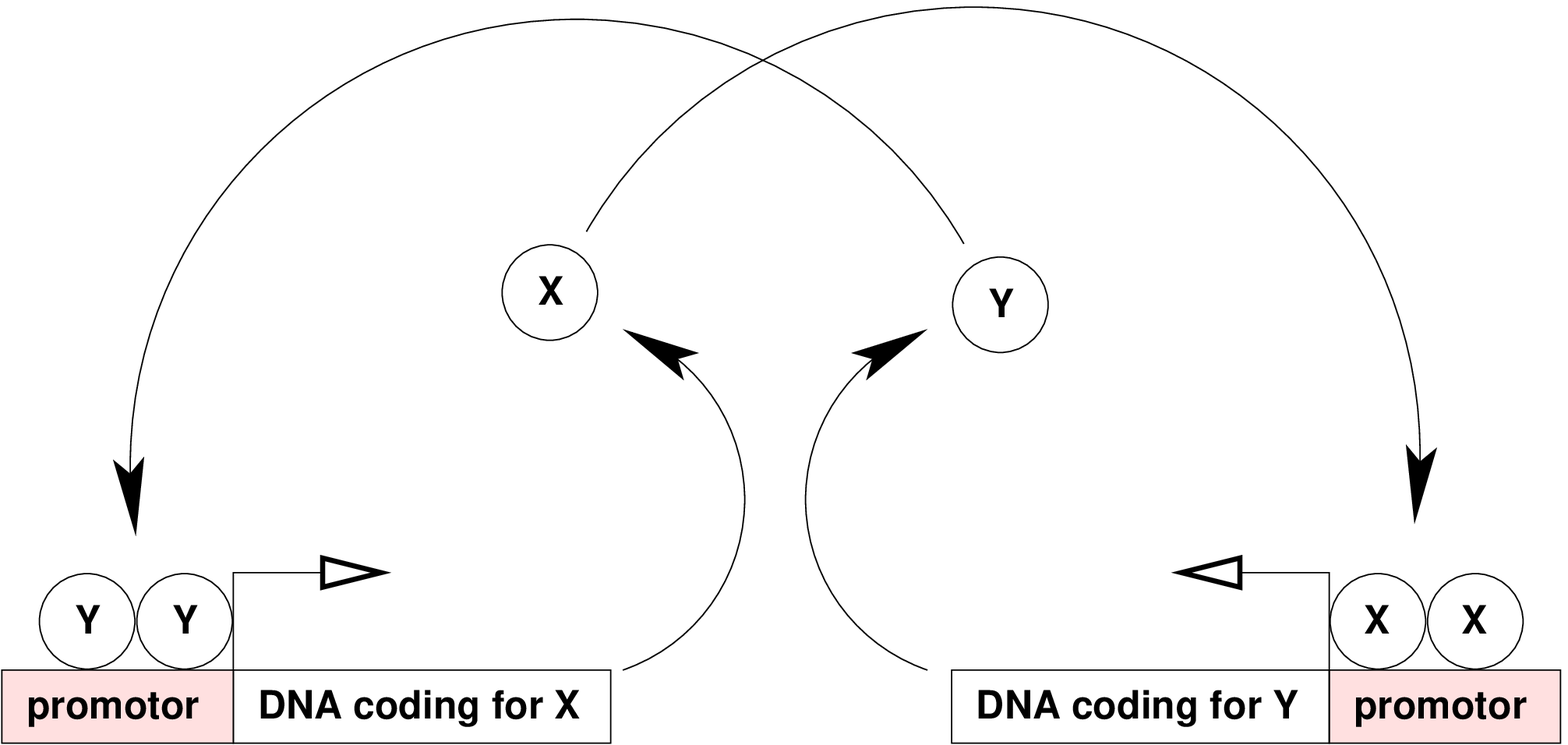}}
\hspace{1cm}
\subfigure[]{\includegraphics[clip,width=5.5cm]{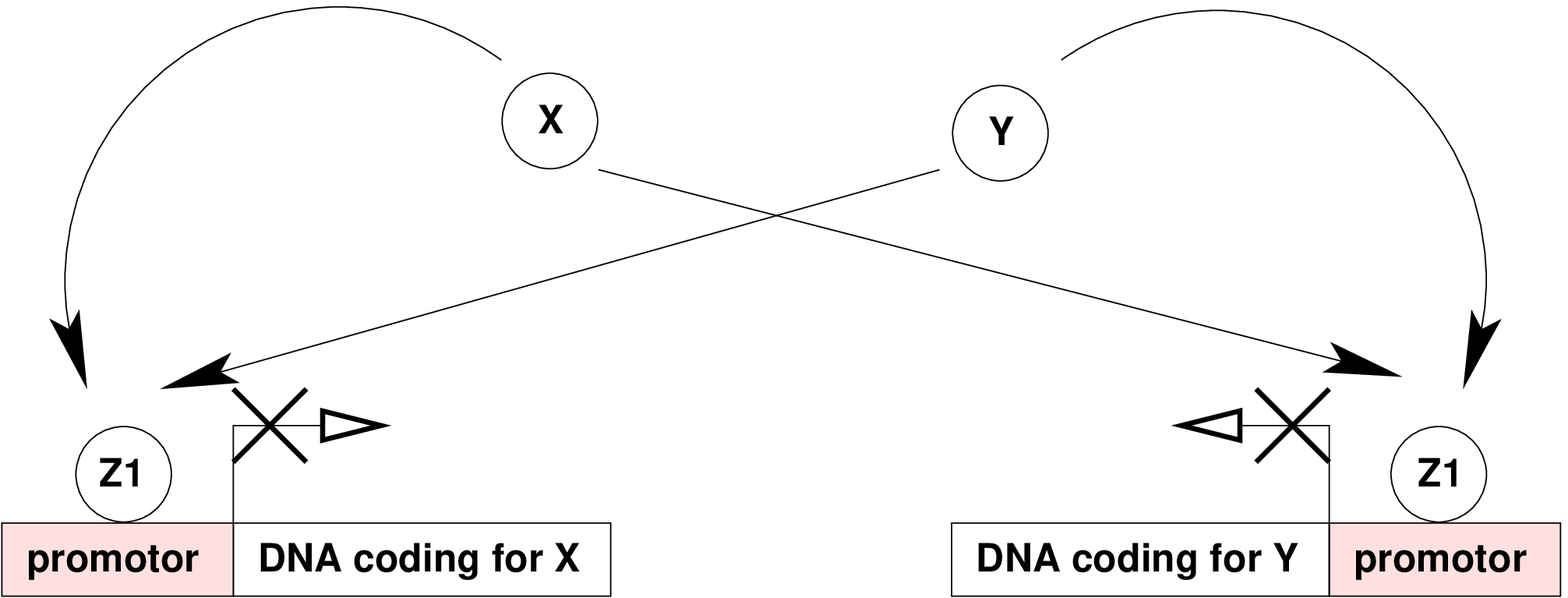}}
\hspace{0.5cm}
\subfigure[]{\includegraphics[clip,width=5.5cm]{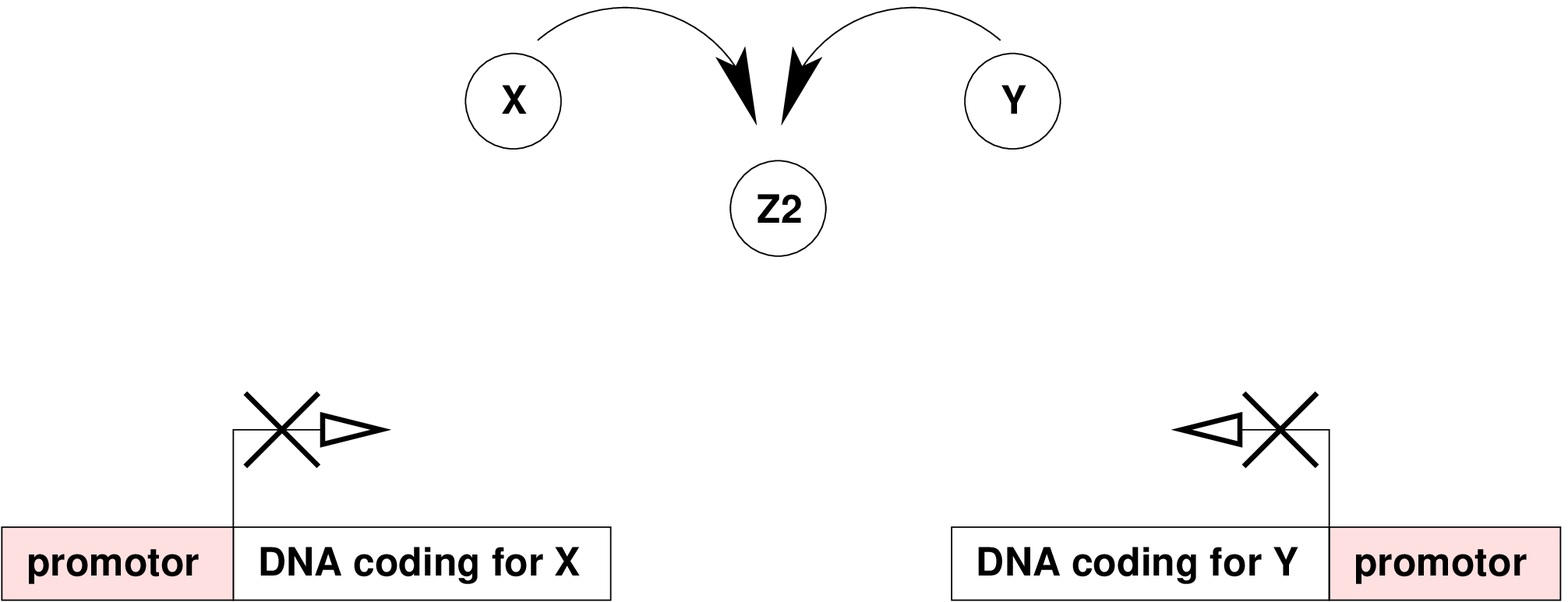}}
\hspace{0.5cm}
\subfigure[]{\includegraphics[clip,width=5.5cm]{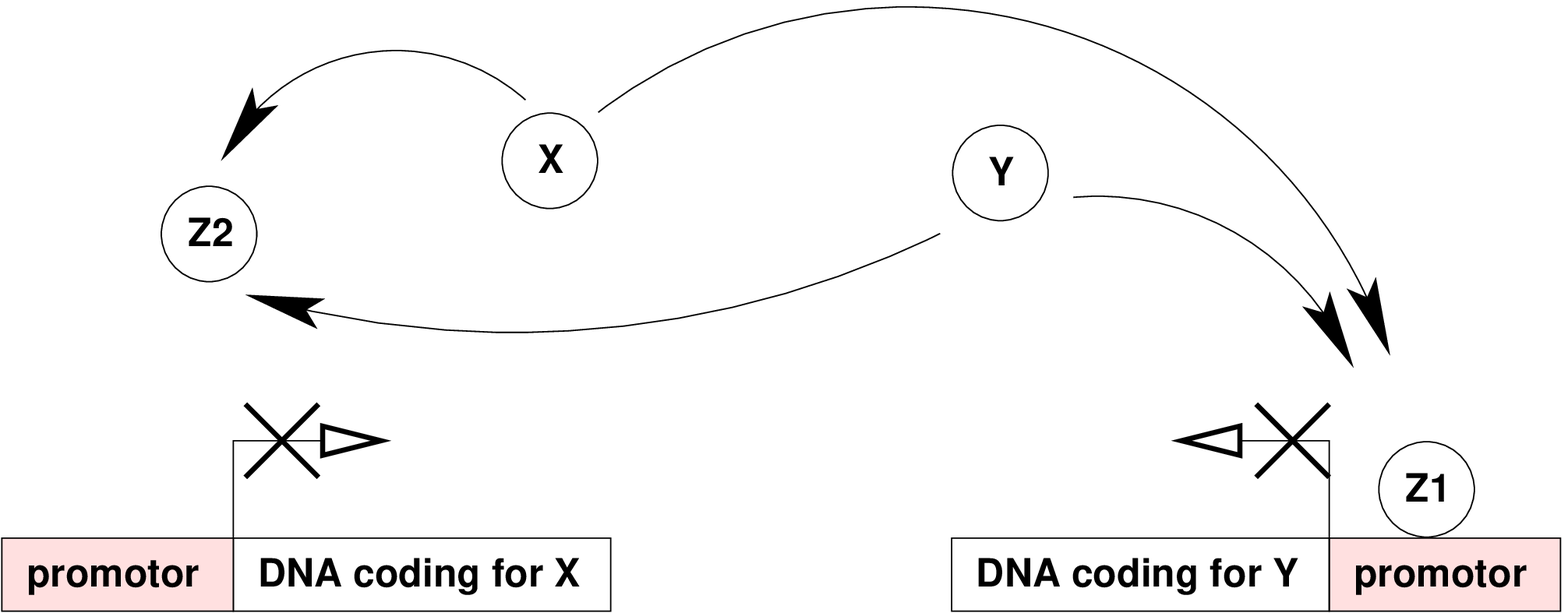}}
\caption[Transcription scenarios]{Principles of transcription
  initiation and inhibition for $X$ and $Y$. (a) Specific
  transcription, i.e.  auto-catalysis by the transcription factor
  itself; (b) Unspecific transcription, i.e. transcription initiated
  by another transcription factor; (c-e) Suggested mechanisms of
  transcription inhibition for $X$ and $Y$ by formation of
  $XY$-complexes: (c) A $XY$-complex, called $Z_1$, bound to the
  promoter regions acts as a repressor; (d) The formation of a
  structurally different $XY$-complex ($Z_2$) competitively inhibits
  the DNA binding of $X$ and $Y$ molecules; (e) Combination of (c) and
  (d) as suggested for GATA-1 and PU.1
  \citep{zhang-p-1999-8705-a,zhang-p-2000-2641-a}}
\label{fig:transcription}
\end{center}
\end{figure}

The general design of the model structure is based on the following
assumptions which are motivated by the experimental observations
  outlined in Section \ref{intro}:

\begin{itemize}
\item Both transcription factors, $X$ and $Y$, are able to act as activator molecules:
  \begin{itemize}
  \item[-] If bound to their own promoter region, $X$ and $Y$ introduce 
    a positive feedback on their own transcription. 
    This process is referred to as \emph{specific} transcription 
    (Fig. \ref{fig:transcription}(a)). 
  \item[-] $X$ and $Y$ are both able to induce an overall transcription 
    which also effects potentially antagonistic transcription factors. 
    Although such an interaction is most likely indirect, 
    for the model we consider a mutual activation of $X$ and $Y$ 
    by the opposing transcription factor, which we refer to 
    as \emph{unspecific} transcription (Fig.  \ref{fig:transcription}(b)).
  \end{itemize}   
We assume that transcription initiation is only achieved by the 
simultaneous binding of two $X$ and $Y$ molecules, respectively 
(i.e., binding cooperativity $c=2$). This assumption is motivated 
by the result that a binding cooperativity $c>1$ is a necessary 
condition for the existence of system bistability 
\citep[see e.g.][]{becskei-a-2001-2528-a,cinquin-o-2005-233-a,gardner-t-2000-339-a}.
\item There is a mutual inhibition of $X$ and $Y$. 
  Within this context, two possible mechanisms, based on the 
  formation of two structurally different complexes of $X$ and $Y$, 
  are considered:
  \begin{itemize}
  \item[-] Joint binding of $X$ and $Y$ molecules to promoter sites
    (Fig.  \ref{fig:transcription}(c)). Here, the DNA-bound
    $XY$-complex ($Z_1$) acts as a transcription repressor, which blocks the
    binding sites. This represents a mode of competition for free
    binding sites.
  \item[-] Formation of another $XY$-complex, called $Z_2$, which
    neither binds to $X$ nor $Y$ DNA binding site (Fig.
    \ref{fig:transcription}(d)). In contrast to $Z_1$, this represents a
    competition for free transcription factor molecules.
  \end{itemize}
Both inhibition mechanisms (including combinations of them) are considered
for $X$ as well as for $Y$.
\end{itemize}

To facilitate the analysis of the mathematical model we make the
following simplifications:
\begin{itemize}
  \item Post-transcriptional regulation is neglected, i.e., the
    transcription of a gene is considered to ultimately result in the
    production of the corresponding protein (here, a transcription
    factor).
  \item Time delays due to transcription and translation processes are
    neglected.
  \item Simultaneous binding of $X$/$Y$ monomers together with a
    $Z_1$-heterodimer, of two $Z_1$-heterodimers, as well as of a $X$
    and a $Y$ monomer at the same promoter are excluded from the
    analysis.
  \item Interactions of $X$, $Y$ as well as the promoter regions of the 
    coding genes with further transcription factors are neglected.
\end{itemize}

Throughout the paper the following notations are used: $x, y$ denote
the molecule concentrations of $X$ and $Y$, respectively. $Z_1$
denotes the DNA bound $XY$-complex and $Z_2$ the structurally
different $XY$- complex, which is not able to bind to promoter DNA.
$D_{x/y}$ denotes free DNA binding sites within the promoter region of
$X$ and $Y$, respectively. In contrast, binding sites occupied by $X$
or $Y$ molecules or by the $XY$-complex $Z_1$ are denoted as
$D_{x/y}^{xx/yy/xy}$.

\subsection{Model equations}\label{model:ode}
With these assumptions one can write down a set of chemical
reaction equations which underly the system dynamics.

The processes of specific and unspecific transcription activation
(see Fig. \ref{fig:transcription}(a),(b)) are described by equations
\eqref{xsprod}-\eqref{xuprod}.
\begin{align}
X+X+D_x\stackrel{K_1}{\rightleftharpoons}D_x^{xx} &; \quad D_x^{xx}\stackrel{s_x}{\rightarrow}D_x^{xx}+X \label{xsprod}\\
Y+Y+D_x\stackrel{K_2}{\rightleftharpoons}D_x^{yy} &; \quad D_x^{yy}\stackrel{u_x}{\rightarrow}D_x^{yy}+X \label{yuprod}\\
Y+Y+D_y\stackrel{K_3}{\rightleftharpoons}D_y^{yy} &; \quad D_y^{yy}\stackrel{s_y}{\rightarrow}D_y^{yy}+Y \label{ysprod}\\
X+X+D_y\stackrel{K_4}{\rightleftharpoons}D_y^{xx} &; \quad D_y^{xx}\stackrel{u_y}{\rightarrow}D_y^{xx}+Y \label{xuprod}
\end{align}

Herein we made the simplifying assumption that the DNA binding of $X$
and $Y$ always occurs as the binding of homodimers.  That means, that
the sequential binding of two monomers, as the second possibility of
DNA binding, is not consider. The process of dimerization, as well as
the DNA binding and dissociation, are regarded to be in quasi steady
state.

Here and throughout the paper, the $K_i={k_{i}}/{\bar{k}_{i}} \;
(i=1,...,7)$ denote the equilibrium (dissociation) constants of the
reactions, with $k_{i}$ and $\bar{k}_{i}$ representing the forward and
backward reaction rate constants, respectively. Finally, it is assumed that
both transcription factor monomers, $X$ and $Y$, decay with first
order kinetics at rates $k_0^x$ and $k_0^y$, whereas dimer-complexes
are assumed to be stable.

The different mutual transcription inhibition mechanisms are
illustrated in Figs. \ref{fig:transcription}(c)-(e). First, we
consider the formation the $XY$-complex $Z_2$ (see Fig.
\ref{fig:transcription}(d))
\be 
X+Y \stackrel{K_5}{\rightleftharpoons} Z_2 \,.
\label{zprod}
\ee 
Under the quasi steady state assumption $Z_2$ does not contribute to the
mathematical description of the system dynamics.

As shown in Fig. \ref{fig:transcription}(c), there is also the
possibility that $X$ and $Y$ form a structurally different heterodimer
$Z_1$, which is able to bind to the promoter regions, acting as a
repressor for $X$ and $Y$ transcription, respectively:
\begin{align}
X+Y+D_x&\stackrel{K_6}{\rightleftharpoons} D_x^{xy}, \label{zrepx} \\
X+Y+D_y&\stackrel{K_7}{\rightleftharpoons} D_y^{xy}. \label{zrepy}
\end{align}
As with the promoter binding of $X$ and $Y$, we collapse dimerization,
which is assumed to be in quasi steady state, and DNA binding into one
process, neglecting the sequential binding of monomers.

Under the posted quasi steady state assumptions, equations
(\ref{xsprod})-(\ref{zrepy}) lead to the following set of ordinary
differential equations:
\begin{align}
\dxt &= -k_{0_x} x + \frac{s_x K_1 x^{2}+u_x K_2 y^2}{1+K_1 x^{2}+K_2 y^{2}+K_6xy} \label{xeq_compl}\\
\dyt &= -k_{0_y} y + \frac{s_y K_3 y^{2}+u_y K_4 x^2}{1+K_3 y^{2}+K_4
  x^{2}+K_7xy} \label{yeq_compl}
\end{align} 
Details of the derivation are given in Appendix \ref{app:TF_dynamics}.

\section{Results}\label{res}

\subsection{Symmetric system}\label{res:symmCase}
To analytically derive steady state as well as potential
bifurcation conditions, we restrict ourself in this section
to the special case of a completely symmetric system, i.e.:
$k_{0_x}=k_{0_y}=k_0$, $s_x=s_y=\ts$, $u_x=u_y=\tu$, $K_1=K_3$,
$K_2=K_4$, and $K_6=K_7$. Using these relations, together with
$\xx=\sqrt{K_1}x$, $\yy=\sqrt{K_1}y$, $k_u={K_2}/{K_1}$,
$k_r={K_6}/{K_1}$, $s ={\sqrt{K_1}}\ts/{k_0}$, $u
={\sqrt{K_1}}\tu/{k_0}$, and $\tau=k_0t$, the system
\eqref{xeq_compl}, \eqref{yeq_compl} can be written in a
dimensionless form as 
\begin{align}
\dxxtt &= -\xx + \frac{s \xx^2 + u k_u \yy^2}{1+\xx^{2}+ k_u \yy^{2}+k_r \xx\yy} \label{xeq_sym},\\
\dyytt &= -\yy + \frac{s \yy^2 + u k_u \xx^2}{1+ k_u \xx^{2}+\yy^{2}+k_r \xx\yy}, \label{yeq_sym} 
\end{align} 

Equations \eqref{xeq_sym} and \eqref{yeq_sym} are a pair of
coupled first order differential equations. The steady state
($\dot\xx=\dot\yy=0$) is defined implicitly by
\begin{align}
\xx &= \frac{s \xx^2 + u k_{u} \yy^2}{1+ \xx^{2} + k_{u} \yy^{2}+ k_{r} \xx\yy}, \label{xeq_sym_equilib}\\
\yy &= \frac{s \yy^2 + u k_{u} \xx^2}{1+ k_{u} \xx^{2} + \yy^{2}+ k_{r} \xx\yy}. \label{yeq_sym_equilib} 
\end{align} 
The domain of these nullclines for $\xx$ and $\yy$ is
restricted by the choice of parameters as outlined in Appendix \ref{app:deformation}.
The intersections of the nullclines correspond to the fixed points
$(\xx^{*},\yy^{*})$ of the differential equations \eqref{xeq_sym} and
\eqref{yeq_sym}. Fixed points on the diagonal $(\xx^*,\xx^*)$ are traced
under the simplifying condition $\xx=\yy$. In this case, equations
\eqref{xeq_sym_equilib} and \eqref{yeq_sym_equilib} can be summarized
by
\be
\xx^{*} = \frac{{\xx^{*}}^2(s + u k_{u})}{1+ {\xx^{*}}^{2}(1 + k_{u} + k_{r})} \; , 
\label{xyeq_sym_equilib}
\ee
The first (trivial) fixed point of equation
\eqref{xyeq_sym_equilib} is $\xx^{*}_{1}=0$. Having eliminated
this solution, the remaining quadratic equation yields two
further non-trivial fixed points at
\be
\xx^{*}_{2/3} = \frac{(s+u k_{u}) \pm \sqrt{(s+u k_{u})^2 - 4(1 + k_{u} + k_{r})}}{2(1 + k_{u} + k_{r})}\, .
\label{xeq_sym_fixPoints}
\ee
$(\xx_2^{*},\xx_2^{*})$ and $(\xx_3^{*},\xx_3^{*})$ are real fixed
points on the diagonal for
\be
s \ge -u k_{u} + 2 \sqrt{1 + k_{u} + k_{r}} \, .
\label{xeq_sym_fixPoints_condition}
\ee
Bifurcation points can be characterized by nullclines intersecting
with equal slopes. The derivatives of equations \eqref{xeq_sym} and
\eqref{yeq_sym} are evaluated to determine explicit conditions for
bifurcation occurrence on the diagonal, considering $s$ as the
bifurcation parameter\footnote{Parameter $s$ is chosen to account for
  changes in the transcriptional activity by enhancer actions or
  modifications in chromatin structure. Furthermore, $s$ is the
  critical parameter that gives rise to the different distinct domains
  for the nullclines as outlined in Appendix \ref{app:deformation}.}.
For simplicity the denominators in equations \eqref{xeq_sym_equilib}
and \eqref{yeq_sym_equilib} are defined as $P_{x}=
1+\xx^{2}+k_{u}\yy^{2}+k_{r}\xx\yy$ and $P_{y}=
1+k_{u}\xx^{2}+\yy^{2}+k_{r}\xx\yy$. The partial derivative of
equation \eqref{xeq_sym_equilib} with respect to $\yy$ leads to
\be
\xx' = \frac{(2s\xx\xx' +2uk_{u}\yy)P_{x} - (s\xx^{2}+uk_{u}\yy^{2})P_{x}'}{P_{x}^{2}}
\label{xeq_sym_derivative}
\ee
with $\xx'={\partial{\xx}}/{\partial{\yy}}$ and
$P_{x}'={\partial{P_{x}}}/{\partial{\yy}}=\xx'(2\xx+k_{r}\yy)+2k_{u}\yy+k_{r}\xx$.
Solving for $\xx'$ yields
\be
\xx' = \frac{2uk_{u}\yy P_{x} -(s\xx^{2}+uk_{u}\yy^{2})(2k_{u}\yy+k_{r}\xx)}
            {P_{x}^{2}-2s\xx P_{x}+(s\xx^{2}+uk_{u}\yy^{2})(2\xx+k_{r}\yy)} \, .
\label{xeq_sym_derivative2}
\ee
For bifurcation points on the diagonal $(\xx=\yy)$, where the
denominators $P_{x}$ and $P_{y}$ simplify to $P^{*}=1+ {\xx^{*}}^{2}(1
+ k_{u} + k_{r})$, equation \eqref{xeq_sym_derivative2} can be
rewritten as
\be
\xx'({P^{*}}^{2}-2s\xx P^{*}+\xx^{3}(s+uk_{u})(2+k_{r})) = 2uk_{u}\xx P^{*} -\xx^{3}(s+uk_{u})(2k_{u}+k_{r}) \, .
\label{xeq_sym_derivative3}
\ee
Inserting $P^{*}$ in the form $P^{*}=\xx(s+uk_{u})$ derived from
equation \eqref{xyeq_sym_equilib} and neglecting the trivial
solution the equality now reads
\be
\xx'(uk_{u}-s+\xx(2+k_{r})) = 2uk_{u}-\xx(2k_{u}+k_{r}) \, .
\label{xeq_sym_derivative4}
\ee

To find the bifurcation points on the diagonal one needs to
study the two distinct cases for $\xx'=1$ and $\xx'=-1$ (see
Appendix \ref{app:bifurcation}).

{\bf Case I ($\xx'=1$):}

\begin{figure}
\begin{center}
\begin{minipage}{.3\textwidth}
\subfigure[]{\includegraphics[clip,width=6cm]{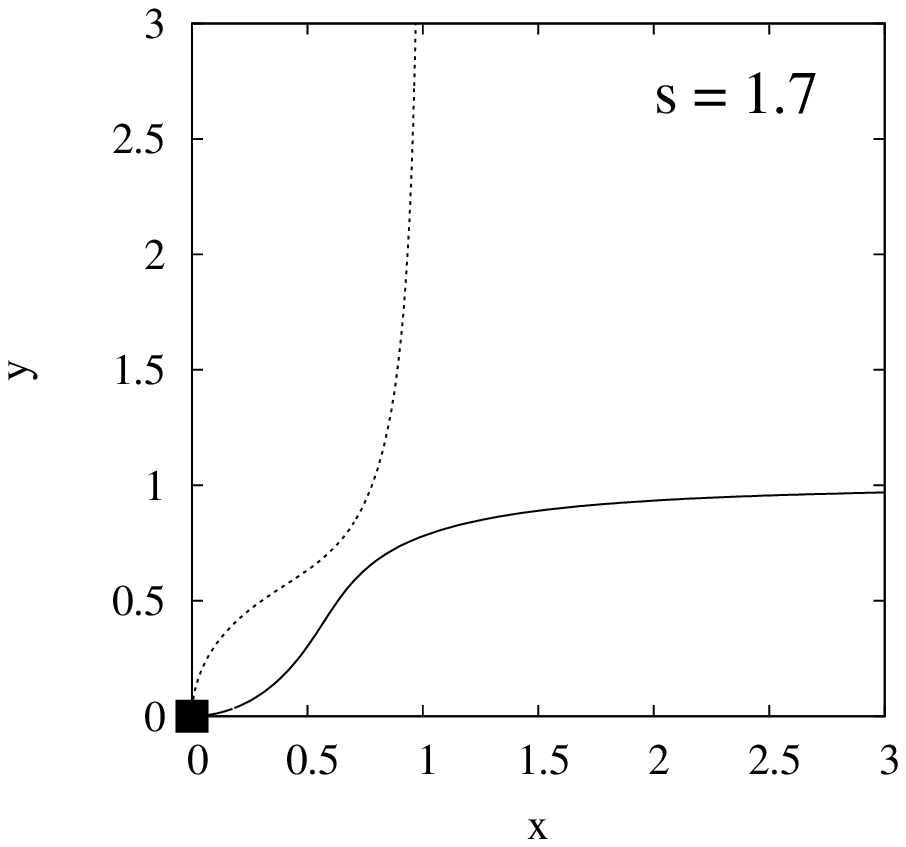}}
\end{minipage}
\begin{minipage}{.3\textwidth}
\subfigure[]{\includegraphics[clip,width=6cm]{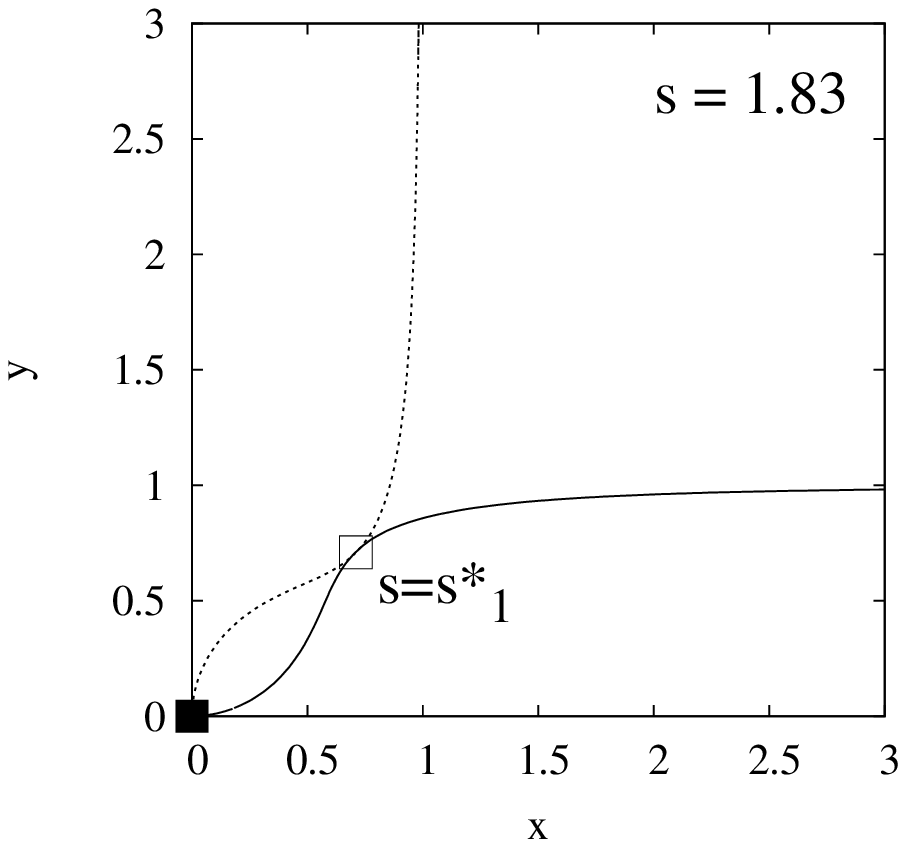}}
\end{minipage}
\begin{minipage}{.3\textwidth}
\subfigure[]{\includegraphics[clip,width=6cm]{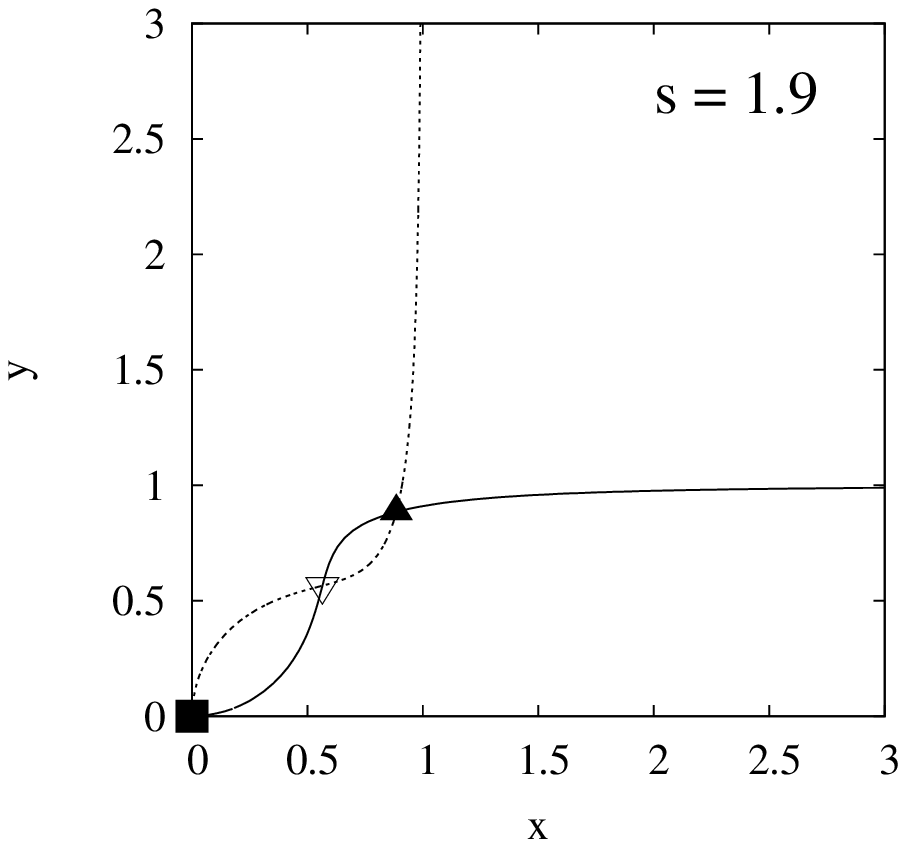}}
\end{minipage}
\begin{center}
\begin{minipage}{.3\textwidth}
\subfigure[]{\includegraphics[clip,width=6cm]{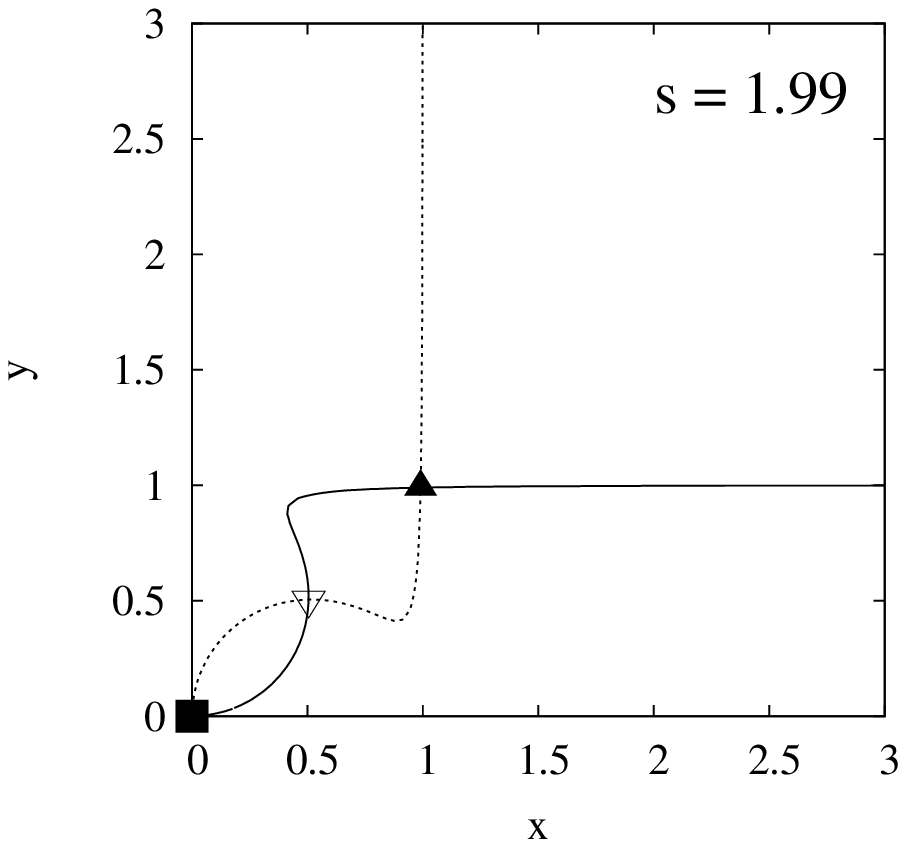}}
\end{minipage}
\begin{minipage}{.3\textwidth}
\subfigure[]{\includegraphics[clip,width=6cm]{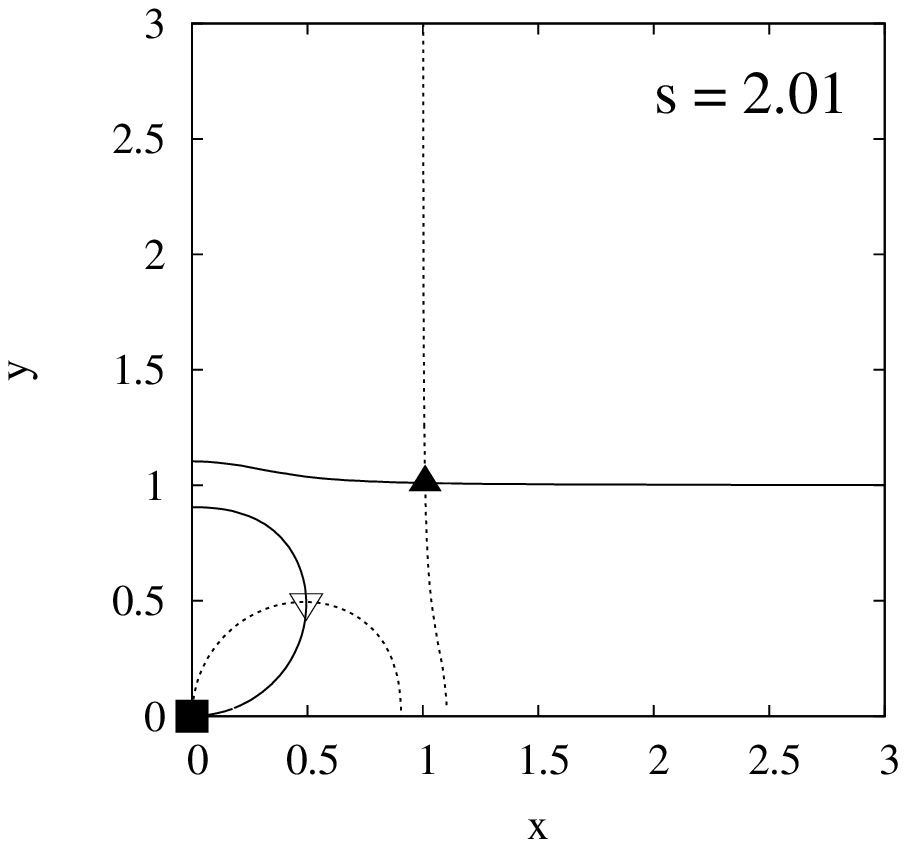}}
\end{minipage}
\end{center}
\begin{minipage}{.3\textwidth}
\subfigure[]{\includegraphics[clip,width=6cm]{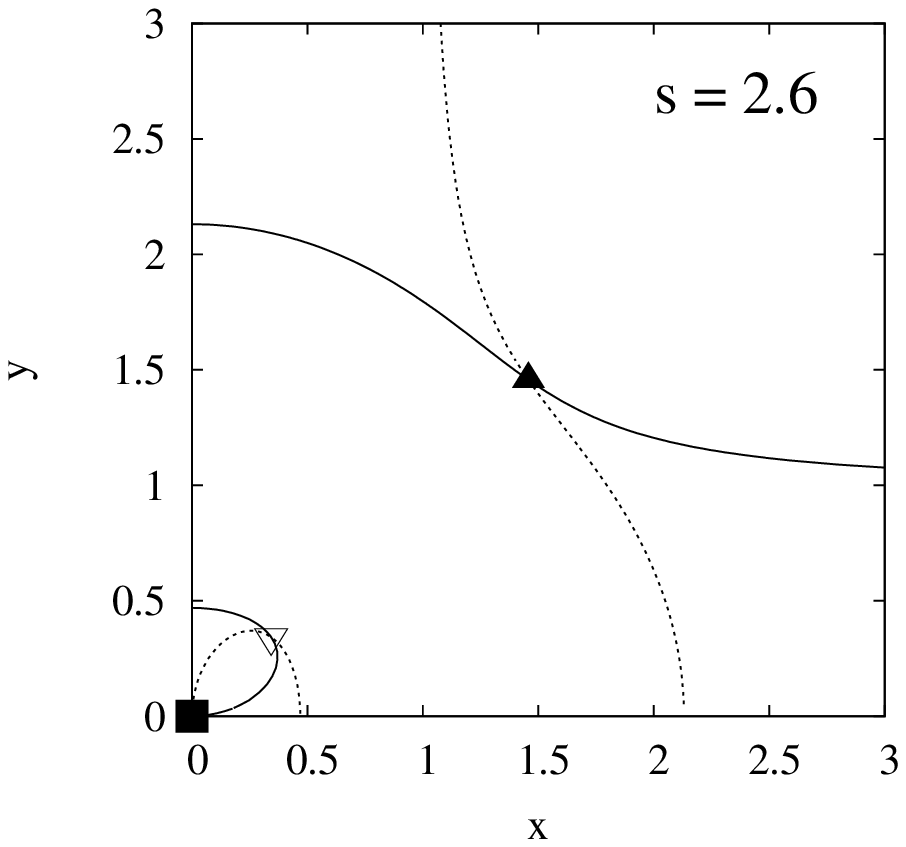}}
\end{minipage}
\begin{minipage}{.3\textwidth}
\subfigure[]{\includegraphics[clip,width=6cm]{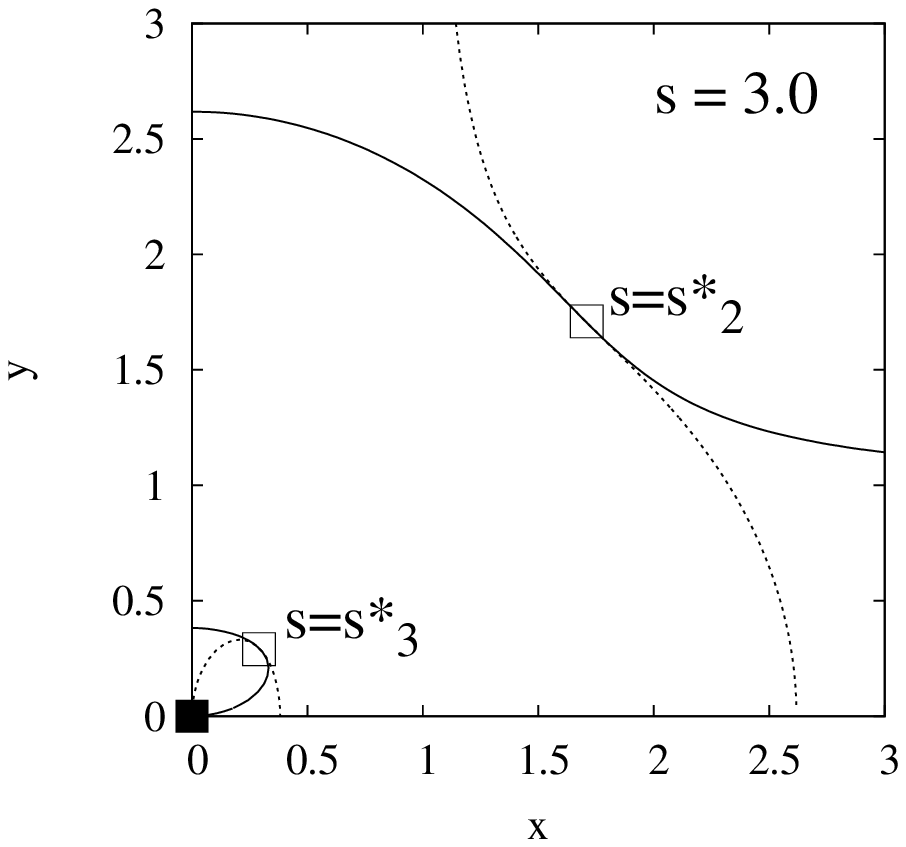}}
\end{minipage}
\begin{minipage}{.3\textwidth}
\subfigure[]{\includegraphics[clip,width=6cm]{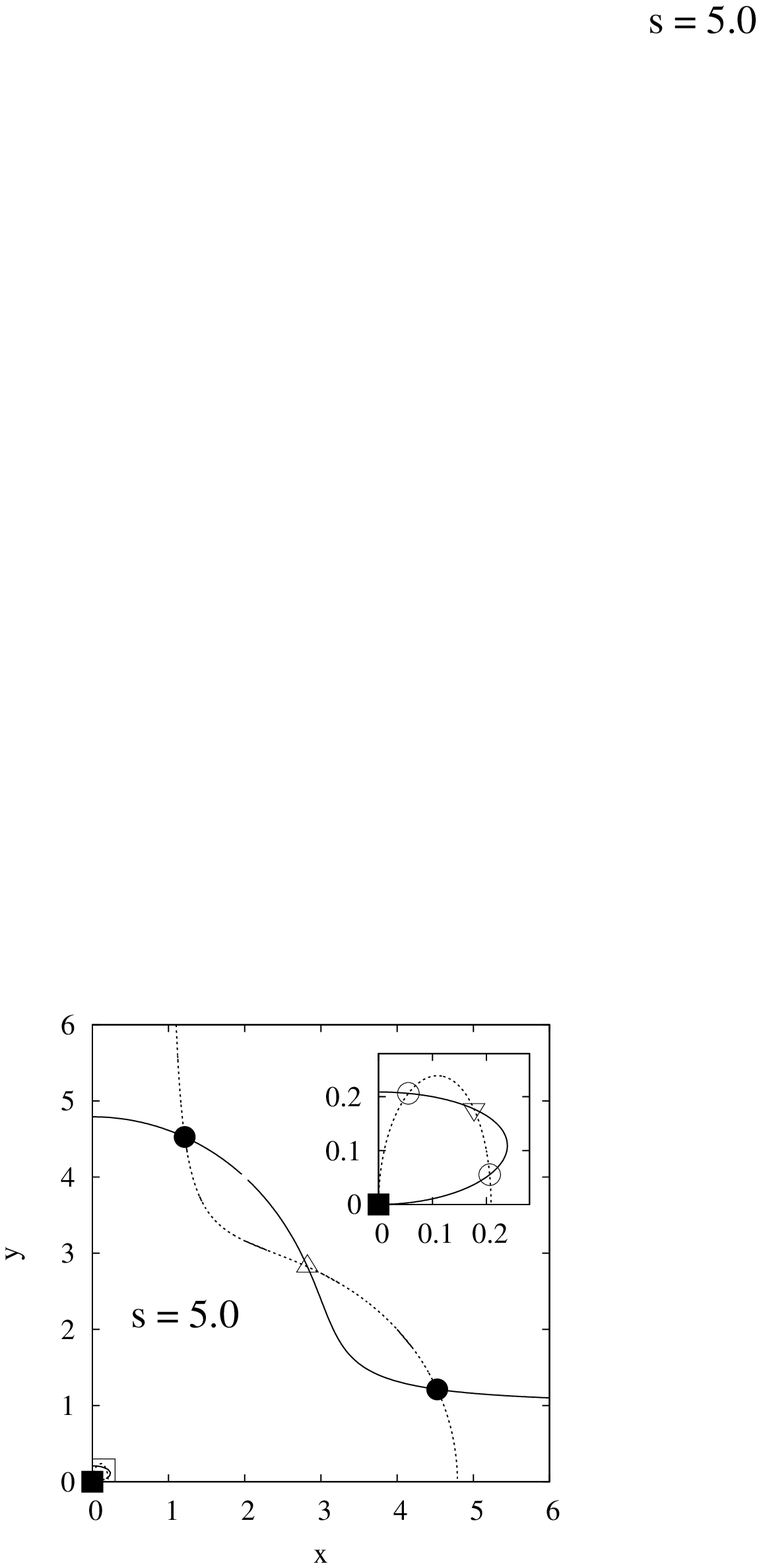}}
\end{minipage}
\caption[Nullclines1]{Deformation of the nullclines for increasing
  values of the bifurcation parameter $s$. The occurrence of the first
  bifurcation at $s_1^{*}$ is depicted in (b), the second bifurcation
  at $s_{1}^{*}=s_{2}^{*}$ in (g). Parameters are $k_{u}=1$, $k_{r}=0$
  and $u=1$. The bifurcation parameter is set to $s=1.7$ (a),
  $s=s_1^{*}=-1+2\sqrt{2} \approx 1.83$ (b), $s=1.9$ (c), $s=1.99$
  (d), $s=2.01$ (e), $s=2.6$ (f), $s=s_{2}^{*}=s_{3}^{*}=3u=3$ (g),
  $s=3.8$ (h). Fixed points are marked as follows: trivial fixed point
  $(0,0)$ - $\blacksquare$, stable/unstable fixed point
  $(\xx_2^{*},\xx_2^{*})$ - $\blacktriangle$/$\triangle$, unstable
  fixed point $(\xx_3^{*},\xx_3^{*})$ - $\triangledown$,
  stable/unstable fixed points off the diagonal - $\bullet$/$\circ$,
  bifurcation point - $\square$.}
\label{fig:ode.nullclines}
\end{center}
\end{figure}

Equation \eqref{xeq_sym_derivative4} satisfies the
condition $\xx'=1$ at
\be
\xx_{\xx'=1} = \frac{s+uk_{u}}{2(1+k_{u}+k_{r})}
\label{xeq_sym_posGrad1}
\ee
This only coincides with the fixed points $\xx^{*}_{2/3}$ derived
in equation \eqref{xeq_sym_fixPoints} if the expression under the
radical in equation \eqref{xeq_sym_fixPoints} vanishes, i.e.,
$\xx^{*}_{2}=\xx^{*}_{3}$. This is true for
\be 
s_1^{*} = -uk_{u} + 2 \sqrt{1+k_{u}+k_{r}}\, ,
\label{xeq_sym_posGrad2} 
\ee
which corresponds to the condition defined in
\eqref{xeq_sym_fixPoints_condition}. This implies that the ``birth" of
the fixed points $(\xx_2^{*},\xx_2^{*})$ and $(\xx_3^{*},\xx_3^{*})$
coincides with the bifurcation condition $\xx'=1$. The sequence of
nullclines in Fig.  \ref{fig:ode.nullclines}(a)-(c) illustrates this
behavior for an unspecific transcription rate $u=1$. The nullclines in
Fig.  \ref{fig:ode.nullclines}(a) do not intersect for $s < s_1^{*}$,
i.e., there is no non-trivial fixed point along the diagonal. For $s =
s_1^{*}$ one common fixed point at $\xx^{*}_{2}(s_1^{*})=
\xx^{*}_{3}(s_1^{*})=({s_1^{*}+uk_{u}})/{2(1+k_{u}+k_{r})}$ exists,
which marks the bifurcation point depicted in Fig.
\ref{fig:ode.nullclines}(b). For $s > s_1^{*}$ two distinct fixed
points $(\xx_2^{*},\xx_2^{*})$ and $(\xx_3^{*},\xx_3^{*})$ exist on
the diagonal, shown in Fig. \ref{fig:ode.nullclines}(c). Whereas the upper
point at $(\xx_2^{*},\xx_2^{*})$ is stable, the lower one at
$(\xx_3^{*},\xx_3^{*})$ is unstable. The nullclines change qualitatively for a
further increase in the bifurcation parameter $s$ as shown in the
sequence Fig. \ref{fig:ode.nullclines}(d),(e).

\begin{figure}
\begin{center}
\begin{minipage}{.3\textwidth}
\subfigure[]{\includegraphics[clip,width=6cm]{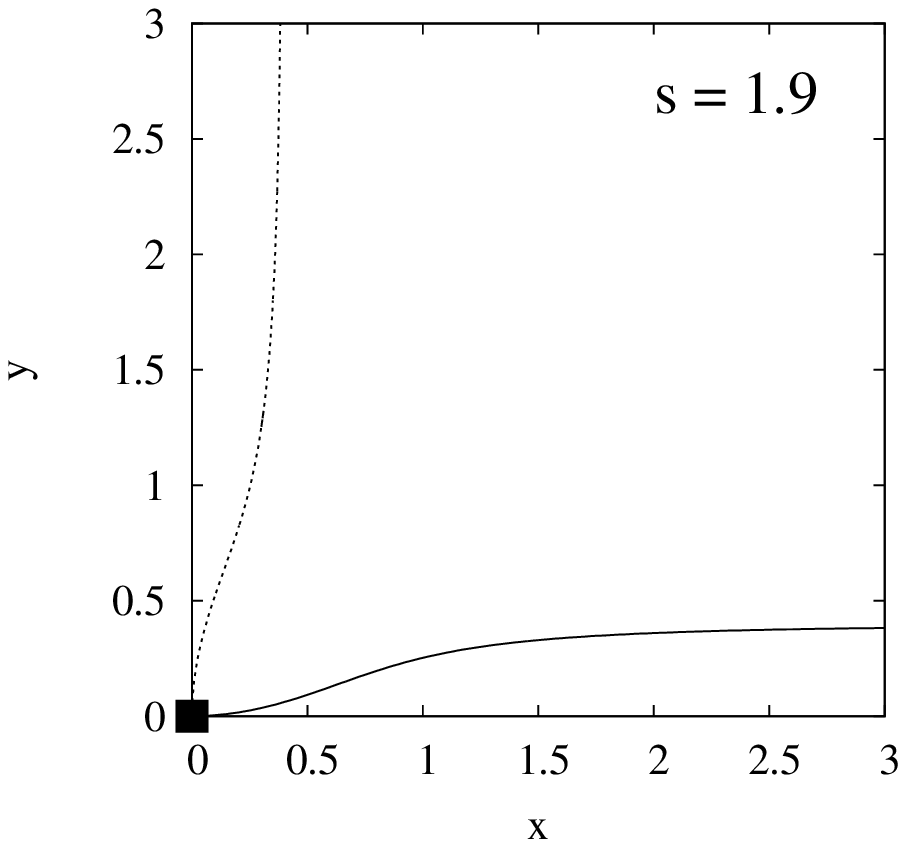}}
\end{minipage}                                 
\begin{minipage}{.3\textwidth}                 
\subfigure[]{\includegraphics[clip,width=6cm]{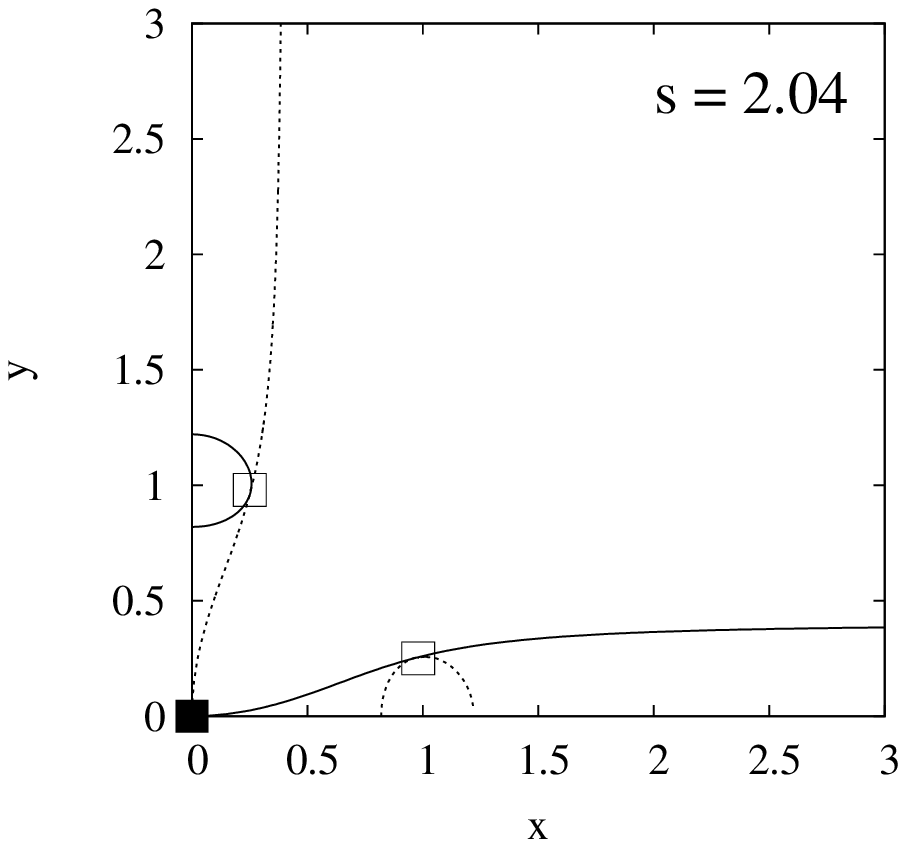}}
\end{minipage}                                 
\begin{minipage}{.3\textwidth}                 
\subfigure[]{\includegraphics[clip,width=6cm]{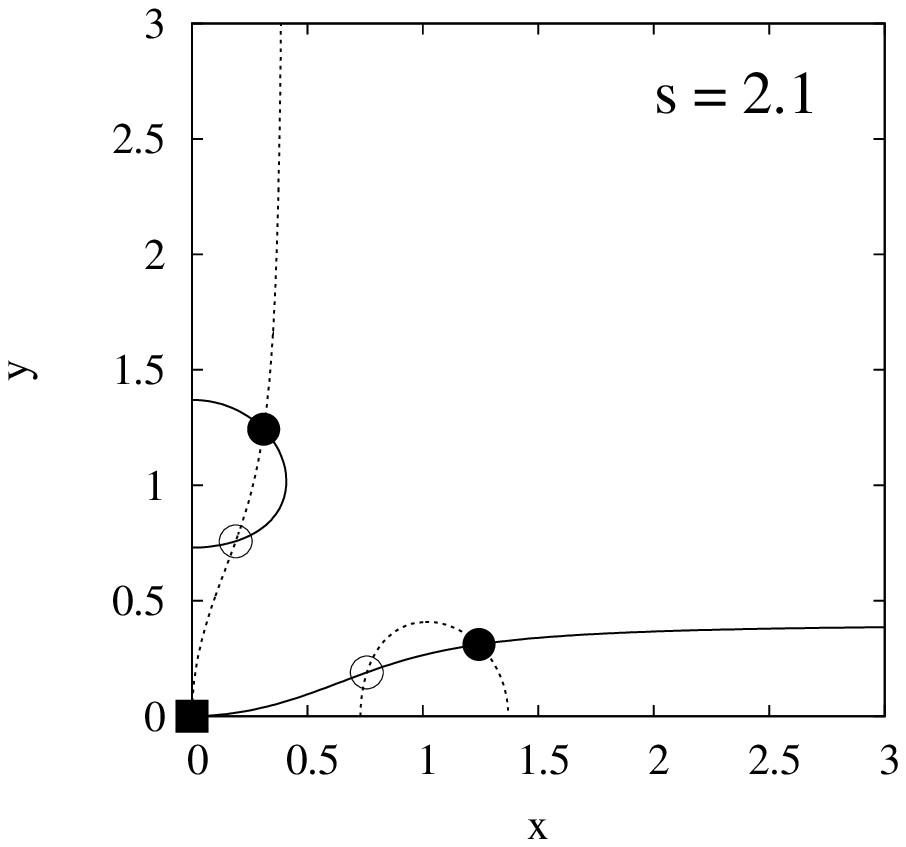}}
\end{minipage}
\begin{minipage}{.3\textwidth}
\subfigure[]{\includegraphics[clip,width=6cm]{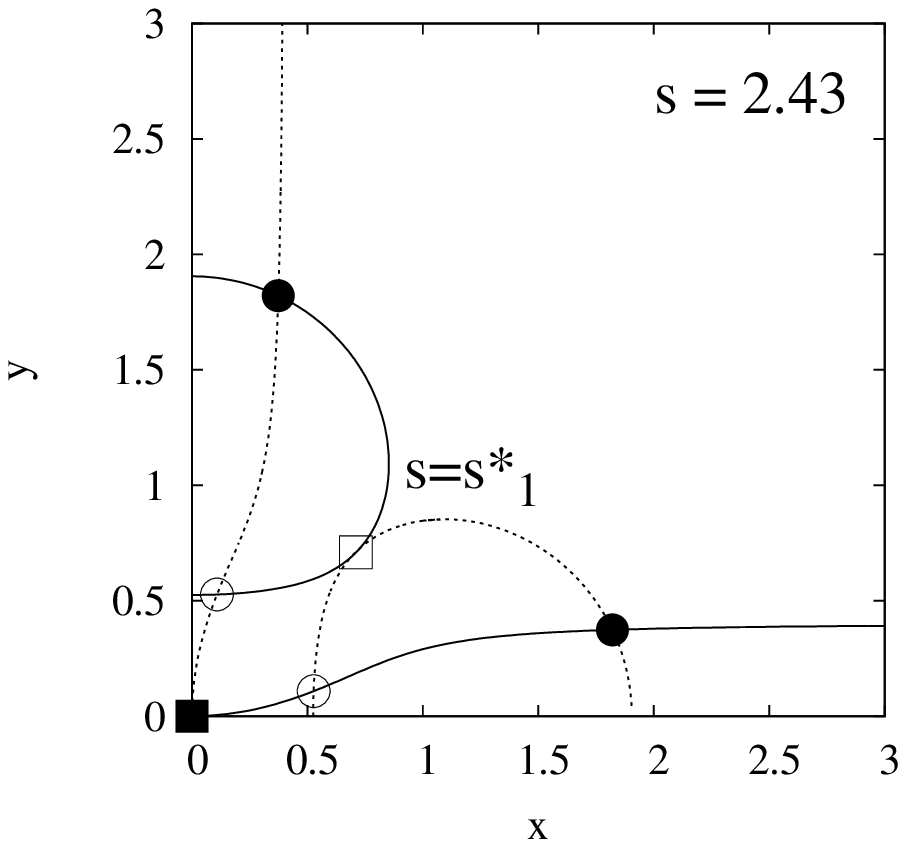}}
\end{minipage}
\begin{minipage}{.3\textwidth}
\subfigure[]{\includegraphics[clip,width=6cm]{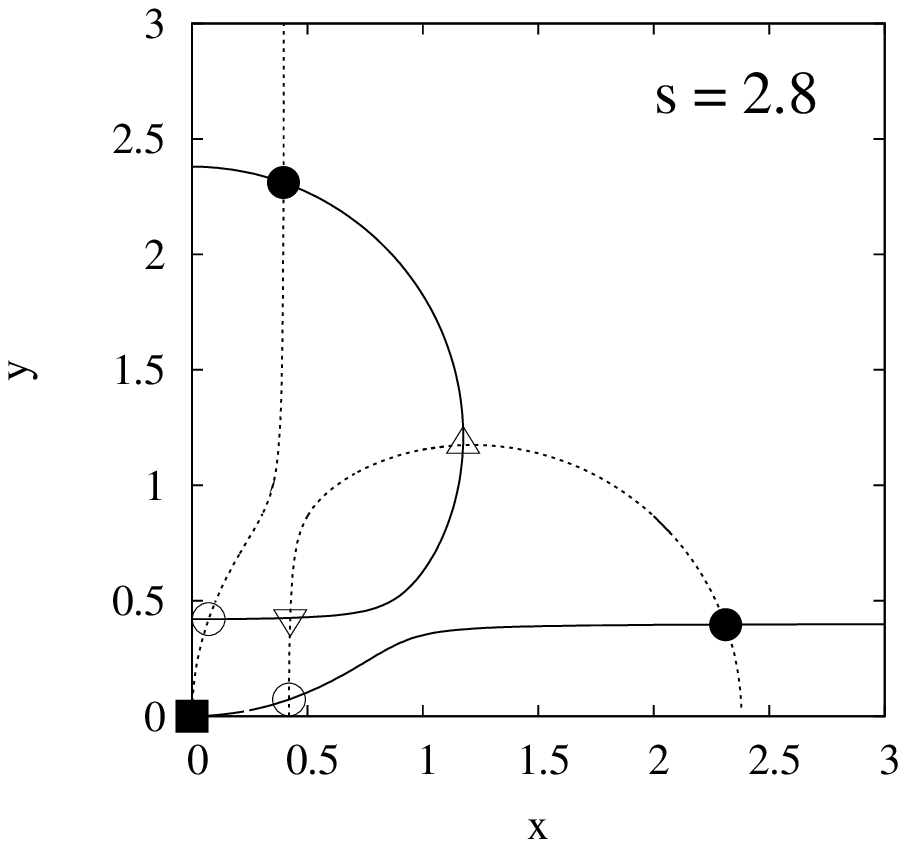}}
\end{minipage}
\begin{minipage}{.3\textwidth}
\subfigure[]{\includegraphics[clip,width=6cm]{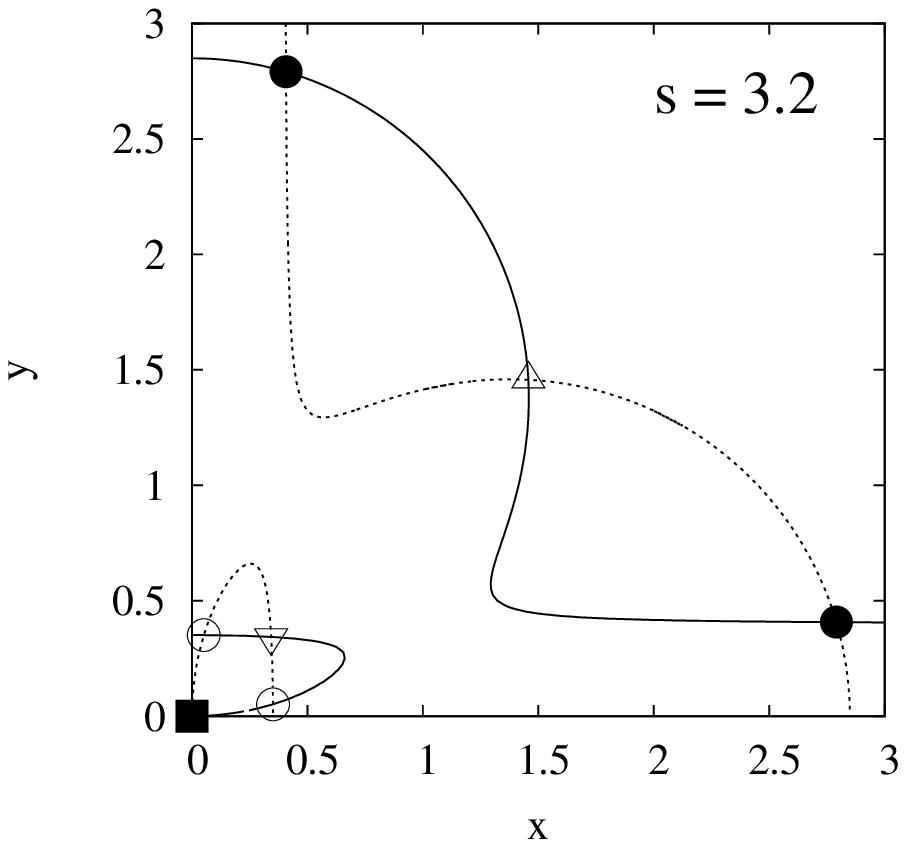}}
\end{minipage}
\caption[Nullclines2]{Deformation of the nullclines for increasing
  values of the bifurcation parameter $s$ for unspecific transcription
  rate $u=0.4$. Parameters are $k_{u}=1$ and $k_{r}=0$. The
  bifurcation parameter is set to $s=1.9$ (a), $s=2.04$ (b), $s=2.1$
  (c), $s=s_{1}^{*} \approx 2.43$ (d), $s=2.8$ (e), $s=3.2$ (f).
  Fixed points are marked according to the caption in Fig.
  \ref{fig:ode.nullclines}.}
\label{fig:ode.nullclines2}
\end{center}
\end{figure}

In the case of a smaller unspecific transcription rate $u=0.4$ the
corresponding bifurcation is illustrated in Fig.
\ref{fig:ode.nullclines2}(d). The two fixed points
$(\xx_2^{*},\xx_2^{*})$ and $(\xx_3^{*},\xx_3^{*})$ generated at
the diagonal are both unstable as depicted in Fig.
\ref{fig:ode.nullclines2}(e),(f).  The qualitative differences between
the scenarios for small and large unspecific transcription $u$ are
more thoroughly investigated in the subsequent paragraphs.

{\bf Case 2 ($\xx'=-1$):}

When $\xx'=-1$ equation \eqref{xeq_sym_derivative4} simplifies to
\be
\xx_{\xx'=-1} = \frac{3uk_{u}-s}{2k_{u}-2}
\label{xeq_sym_negGrad}
\ee
Equating $\xx_{\xx'=-1}=\xx^{*}_{2/3}$ from equation
\eqref{xeq_sym_fixPoints} leads to a dependency on the parameters
$s$, $u$, $k_{u}$ and $k_{r}$. Two further bifurcation points are
obtained at:
\begin{align}
s_{2}^{*} &= \frac{u k_{u}(1+3k_{r}+5k_{u}) + 
2(k_{u}-1)\sqrt{1-k_{r}-3k_{u}+4k_{u}^{2}u^{2}}}{(-1+k_{r}+3k_{u})}
\label{xeq_sym_negGrad1}\\
s_{3}^{*} &= \frac{u k_{u}(1+3k_{r}+5k_{u}) -
2(k_{u}-1)\sqrt{1-k_{r}-3k_{u}+4k_{u}^{2}u^{2}}}{(-1+k_{r}+3k_{u})}
\label{xeq_sym_negGrad2}
\end{align}
To guarantee the existence of these bifurcations at the
diagonal, $s_{2/3}^{*} \geq s_1^{*}$ is required. The case
$s_{2/3}^{*}<s_1^{*}$ indicates that bifurcations occur off
the diagonal.

For the special case $k_{u}=1$ the conditions for the occurrence of
these bifurcations simplify to $s_{2}^{*}=s_{3}^{*}=3u$ (given
$s_{2}^{*}=s_{3}^{*}>s_1^{*}$ which is true for $u> 1/\sqrt{2}$). Since
this condition is valid for both fixed points, $(\xx_2^{*},\xx_2^{*})$
and $(\xx_3^{*},\xx_3^{*})$, it indicates that the bifurcations occur
at the same bifurcation parameter $s=3u$. Fig.
\ref{fig:ode.nullclines}(f)-(h) depicts the bifurcations for
both fixed points in the case $u=1$, $k_{u}=1$. After a deformation of
the nullclines the intersections in Fig.  \ref{fig:ode.nullclines}(f)
still represent the fixed points $(\xx_2^{*},\xx_2^{*})$ and
$(\xx_3^{*},\xx_3^{*})$ for $s<s_{2}^{*}=s_{3}^{*}$. In Fig.
\ref{fig:ode.nullclines}(g) the nullclines for $s=s_{2}^{*}=s_{3}^{*}$
intersect with the same local slope at $\xx^{*}_{2}$ as well as at
$\xx^{*}_{3}$. This marks the bifurcation point for both fixed points,
that coincides for $k_{u}=1$.  Fig. \ref{fig:ode.nullclines}(h)
illustrates the new fixed points off the diagonal, which are stable
bifurcating from $(\xx_2^{*},\xx_2^{*})$ and unstable bifurcating from
$(\xx_3^{*},\xx_3^{*})$. The fixed point $(\xx_2^{*},\xx_2^{*})$
itself changes the stability and becomes unstable,
$(\xx_3^{*},\xx_3^{*})$ remains unstable as before.

For small $u$ the condition for the occurrence of further bifurcations
$s_{2/3}^{*} \geq s_1^{*}$ is violated.  Numerical results indicate
that two saddle-node bifurcations form fixed points off the diagonal
at $s<s_1^{*}$ as depicted in the sequence of nullclines in Figs.
\ref{fig:ode.nullclines2}(b),(c). The saddle-node bifurcation on the
diagonal is observed at $s_1^{*}$.  For large $s$ these scenarios show
a comparable pattern of two up-regulated steady states with one high
and one low expressed component and a further stable fixed point at
$(0,0)$ (compare Figs. \ref{fig:ode.nullclines}(h) and
\ref{fig:ode.nullclines2}(f)).

\begin{figure}
\begin{center}
\begin{minipage}{.49\textwidth}
\subfigure[]{\includegraphics[clip,width=7.2cm]{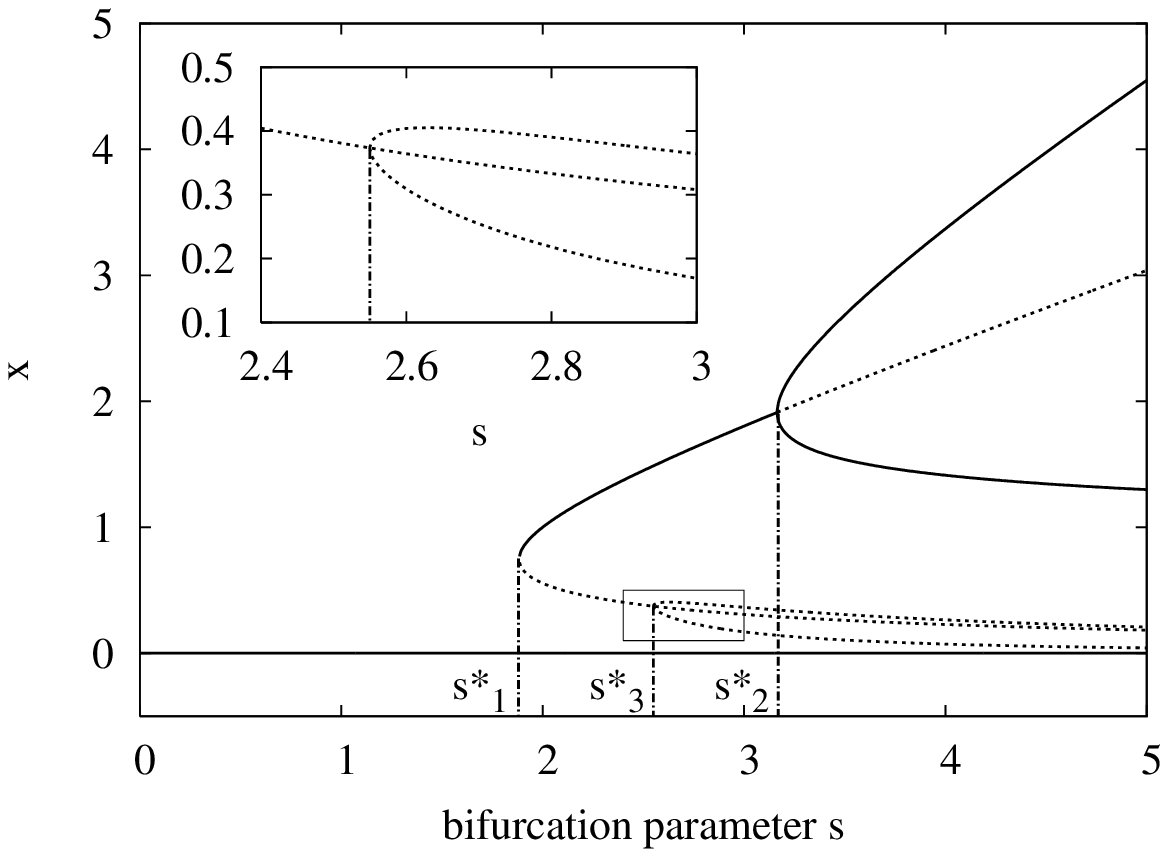}}
\end{minipage}
\begin{minipage}{.49\textwidth}
\subfigure[]{\includegraphics[clip,width=7.2cm]{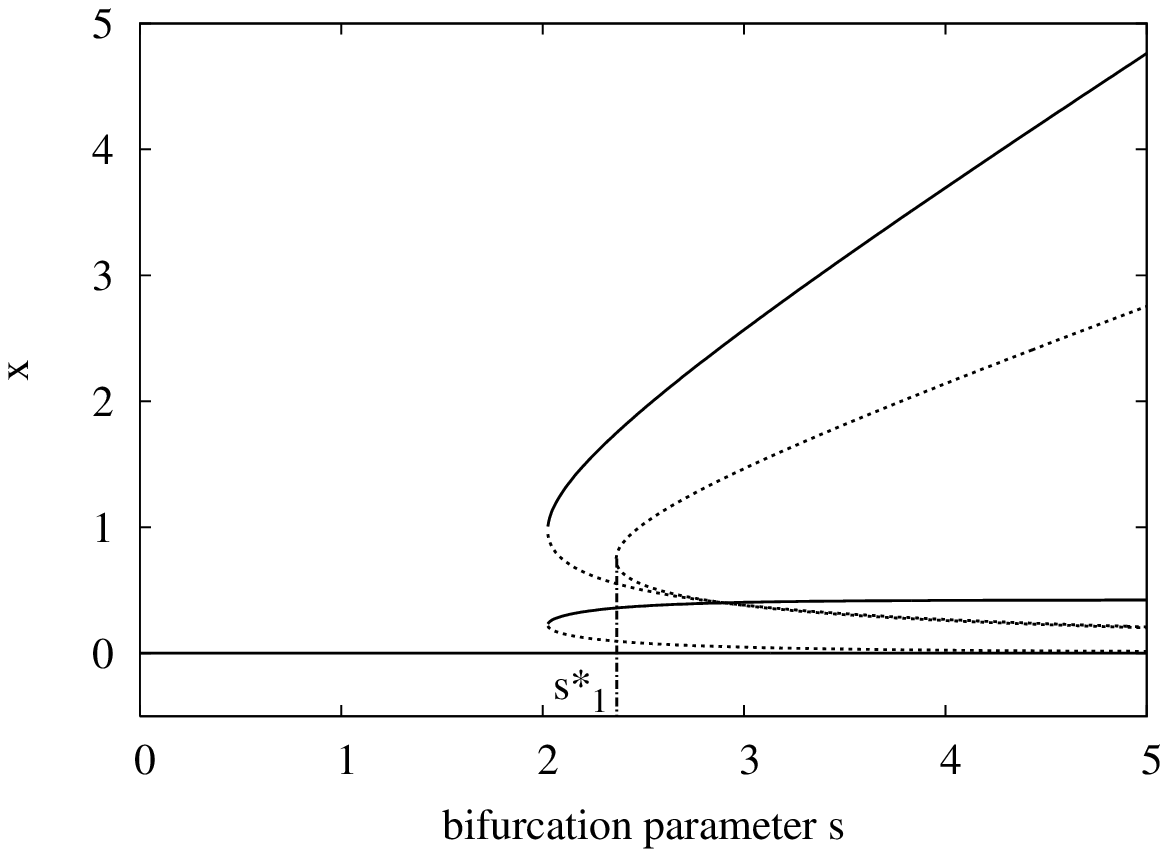}}
\end{minipage}
\caption[Bifurcation diagram]{Bifurcation diagrams $x$ vs. $s$ with
  $k_{u}=0.8$, $k_{r}=0$, $u=1$ (a) and $0=0.4$ (b). The stability of
  the steady states is coded as follows: solid line - stable, dashed
  line - unstable.}
\label{fig:ode.bifurcDiagram}
\end{center}
\end{figure}

The bifurcation diagrams in Fig. \ref{fig:ode.bifurcDiagram}
comprise the above findings for $k_u=0.8$. The $\xx$-coordinate for
the fixed points is shown depending on the bifurcation parameter $s$.
For $u=1$ (Fig.  \ref{fig:ode.bifurcDiagram}(a)) the birth of two
fixed points through a saddle-node bifurcation can be seen at
$s_1^{*}$, given by equation \eqref{xeq_sym_posGrad2}.  Condition
\eqref{xeq_sym_negGrad1} defines the occurrence of the pitchfork
bifurcation on the upper branch $(\xx_2^{*},\xx_2^{*})$ at
$s_{2}^{*}$, whereas condition \eqref{xeq_sym_negGrad2} is the
equivalent for the lower branch $(\xx_3^{*},\xx_3^{*})$ at
$s_{3}^{*}$. Note that the additional condition $s_{2/3}^{*} \geq
s_1^{*}$ is fulfilled. The upper branch gives rise to three fixed
points, one unstable (arising from the existing stable fixed point) at
the diagonal at $\xx^{*}_{2}$ and two new stable fixed points
branching off this axis. For the lower case all three fixed points are
unstable for $s>s_{3}^{*}$. The inset in Fig.
\ref{fig:ode.bifurcDiagram}(a) enlarges this bifurcation occurring at
$s_{3}^{*}$. Fig. \ref{fig:ode.bifurcDiagram}(b) illustrates the
equivalent scenario for $u=0.4$. The saddle-node bifurcation at
$s_1^{*}$ represents the formation of the unstable fixed points
$(\xx_2^{*},\xx_2^{*})$ and $(\xx_3^{*},\xx_3^{*})$. In addition, two
further saddle node bifurcations exist that also form stable fixed
points. Since $s_{2/3}^{*} < s_1^{*}$ these bifurcations do not occur
on the diagonal. All branches in Fig.  \ref{fig:ode.bifurcDiagram}
that do not represent fixed points on the diagonal were determined
numerically.

\begin{figure}
\begin{center}
\subfigure[]{\includegraphics[clip,width=12cm]{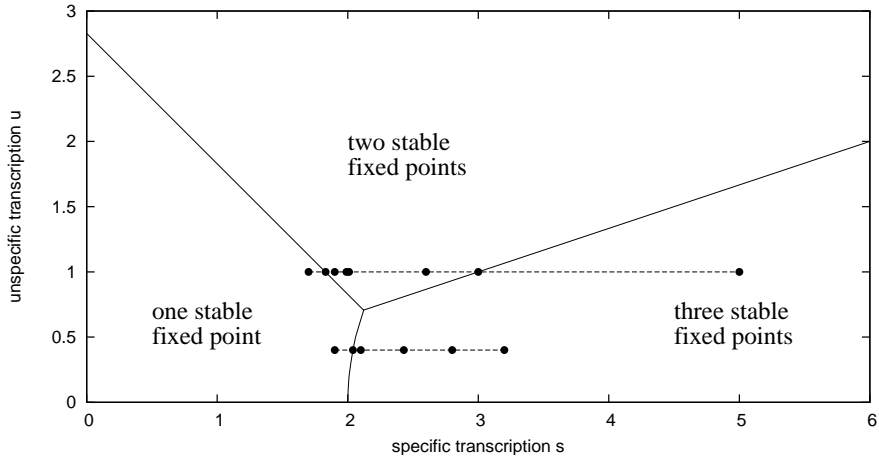}}
\caption[Phase space]{Phase space diagram $u$ vs. $s$ with $k_{u}=1$,
  $k_{r}=0$. The lines separating the distinct regions of
  multi-stability are determined by equations
  \eqref{xeq_sym_posGrad2}, \eqref{xeq_sym_negGrad1} and, for the
  lower branch, numerical results. In the lower left region only one
  stable fixed point at $(0,0)$ exists. In the region marked with
  ``two stable fixed points'' one additional up-regulated stable fixed
  point exists besides the one at $(0,0)$. In ``three stable fixed
  points'' region two additional up-regulated stable fixed points
  exist. The dashed horizontal lines correspond to the sequences of
  nullclines in Figs.  \ref{fig:ode.nullclines} and
  \ref{fig:ode.nullclines2}.}
\label{fig:ode.phaseSpace}
\end{center}
\end{figure}

Fig. \ref{fig:ode.phaseSpace} provides an overview of regions of
multi-stability in the phase space $u$ vs. $s$. Distinct regions with
different numbers of stable steady states are identified depending on
the combination of the dimensionless parameters. Lines of separation
are determined by equations \eqref{xeq_sym_posGrad2},
\eqref{xeq_sym_negGrad1} and, for the lower branch, numerical results.
The sequence of nullclines given in Fig.  \ref{fig:ode.nullclines} is
illustrated by the dashed line at $u=1$, with the dots referring to
the subfigures for varying $s$.  The dashed line at $u=0.4$ gives a
similar representation, with its correspondence in Fig.
\ref{fig:ode.nullclines2}.

Figs. \ref{fig:ode.nullclines} and \ref{fig:ode.nullclines2} both
indicate that the basin of attraction for the fixed point at the
origin $(0,0)$ is separated from the basins of attraction of the
up-regulated stable states by a set of unstable fixed points. The
sequences of graphs also illustrate that these unstable fixed points
move towards the fixed point at $(0,0)$ for increasing $s$, thus
continously reducing the size of its basin of attraction. However,
this size characterizes the stability of the fixed point at the origin
$(0,0)$ in response to external perturbations. Unlike the intermediate
stable steady state, arising from the bifurcation at $s_1^{*}$
depicted in Fig.  \ref{fig:ode.bifurcDiagram}(a), where a dynamically
increasing $s$ inevitably leads to one of the two up-regulated fixed
points, the escape from the fixed point $(0,0)$ needs to be triggered
by a perturbation that exceeds the size of its basin of attraction.
Given the position of the unstable fixed point at the diagonal
$(\xx_3^{*},\xx_3^{*})$ as a function of $s$ in equation
\eqref{xeq_sym_fixPoints}, an appropriate measure for the size of the
basin of attraction is provided.

\subsection{Asymmetric system}\label{res:asymm}

As indicated by \citet{zhang-p-1999-8705-a, zhang-p-2000-2641-a}
the inhibition of PU.1 by GATA-1 and the converse are based on
different mechanisms. The formation of the PU.1-GATA-1 complex,
which we refer to as a $Z_2$-complex, prevents free transcription
factors from binding to their specific DNA binding sites. A
competitive inhibition in this form affects both transcription
factors, although \citet{ zhang-p-2000-2641-a} do not explicitly outline the
consequences of binding of the PU.1-GATA-1 complex to the PU.1
binding site. On the other hand, GATA-1 prevents the binding of
c-Jun to the DNA bound PU.1 protein and thus disables the
transcription initiation of PU.1. This process explicitly targets
the PU.1 binding sites and introduces a functional asymmetry of
inhibition mechanisms.

The mathematical counterpart of this  asymmetry is a specific
binding rate $K_6>0$ while keeping $K_7=0$ (see equations
(\ref{xeq_compl}), (\ref{yeq_compl})). In terms of the
dimensionless formulation in equations \eqref{xeq_sym} and
\eqref{yeq_sym} this translates into two different rate constants
$k_{r_{x}}>0$ and $k_{r_{y}}=0$. The additional binding mode
($k_{r_x}>0$) can be interpreted as a reduction in the
transcriptional activity of the $X$ gene conferring a disadvantage
relative to $Y$.

\begin{figure}
\begin{center}
\begin{minipage}{.33\textwidth}
\subfigure[]{\includegraphics[clip,width=6cm]{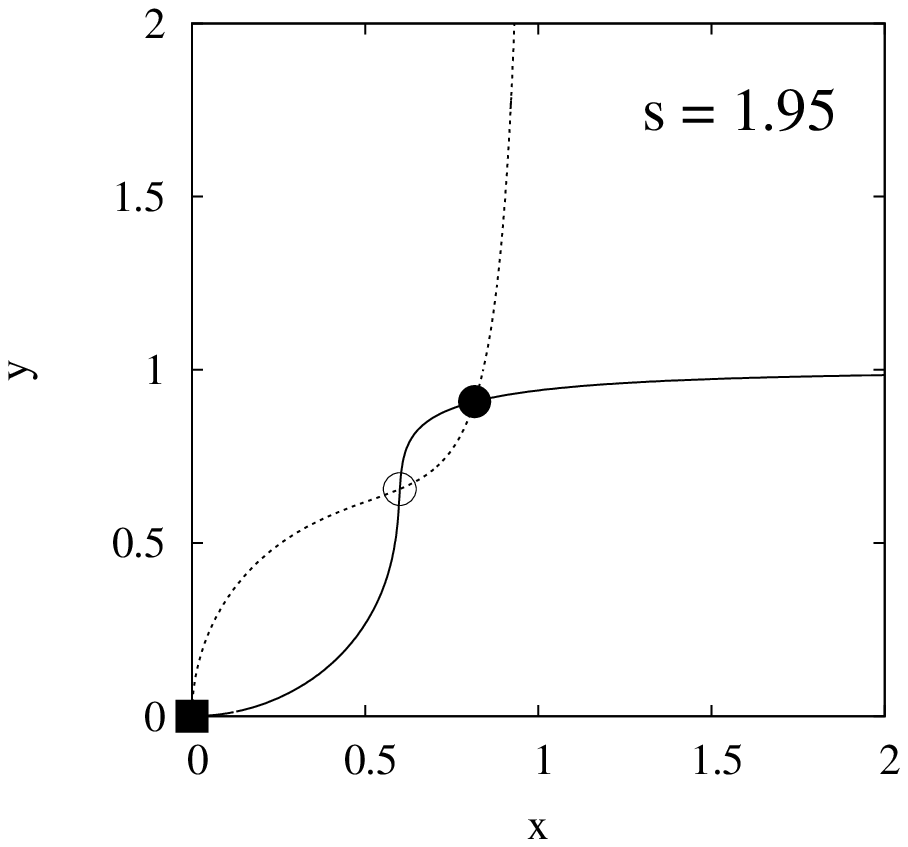}}
\end{minipage}
\begin{minipage}{.33\textwidth}
\subfigure[]{\includegraphics[clip,width=6cm]{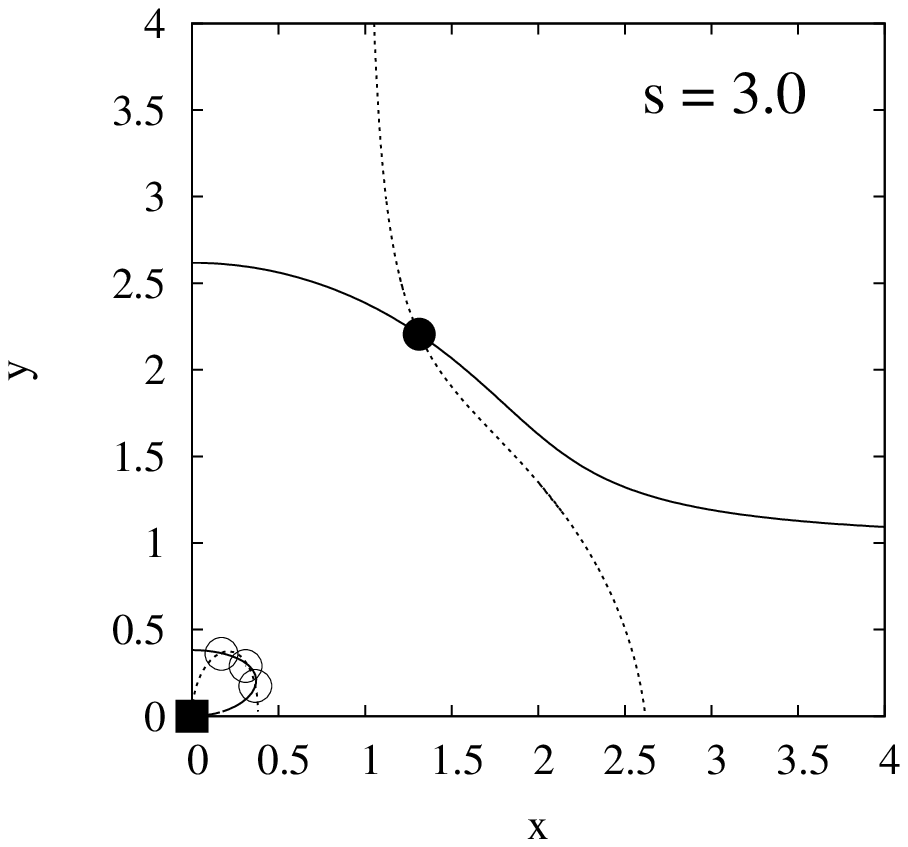}}
\end{minipage}
\begin{minipage}{.32\textwidth}
\subfigure[]{\includegraphics[clip,width=6cm]{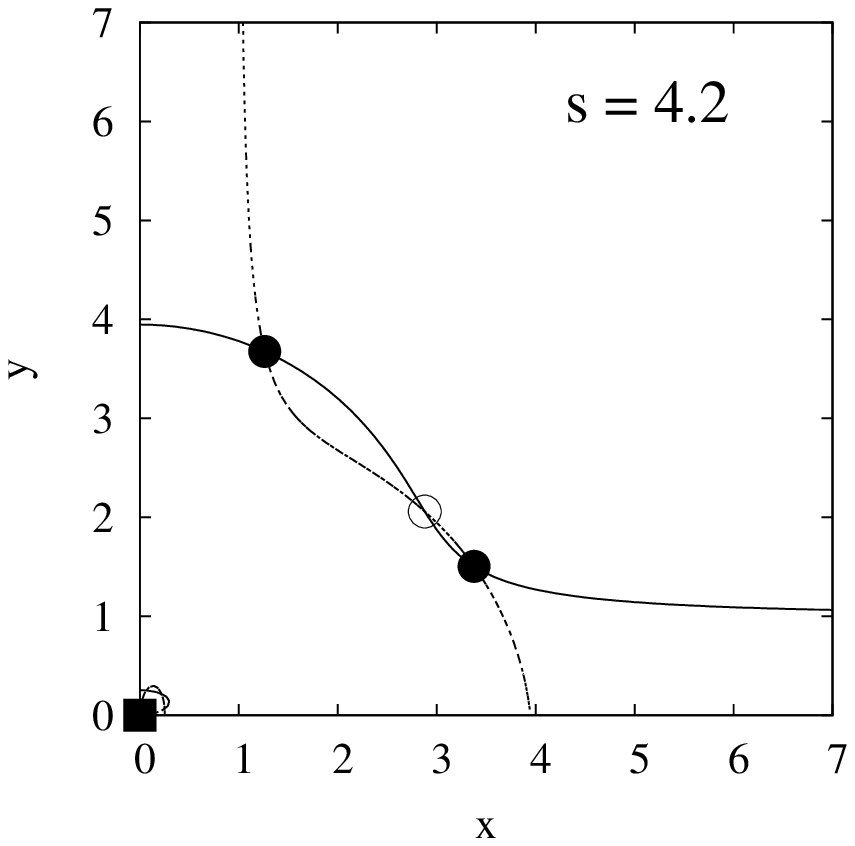}}
\end{minipage}
\begin{minipage}{.32\textwidth}
\subfigure[]{\includegraphics[clip,width=5.0cm]{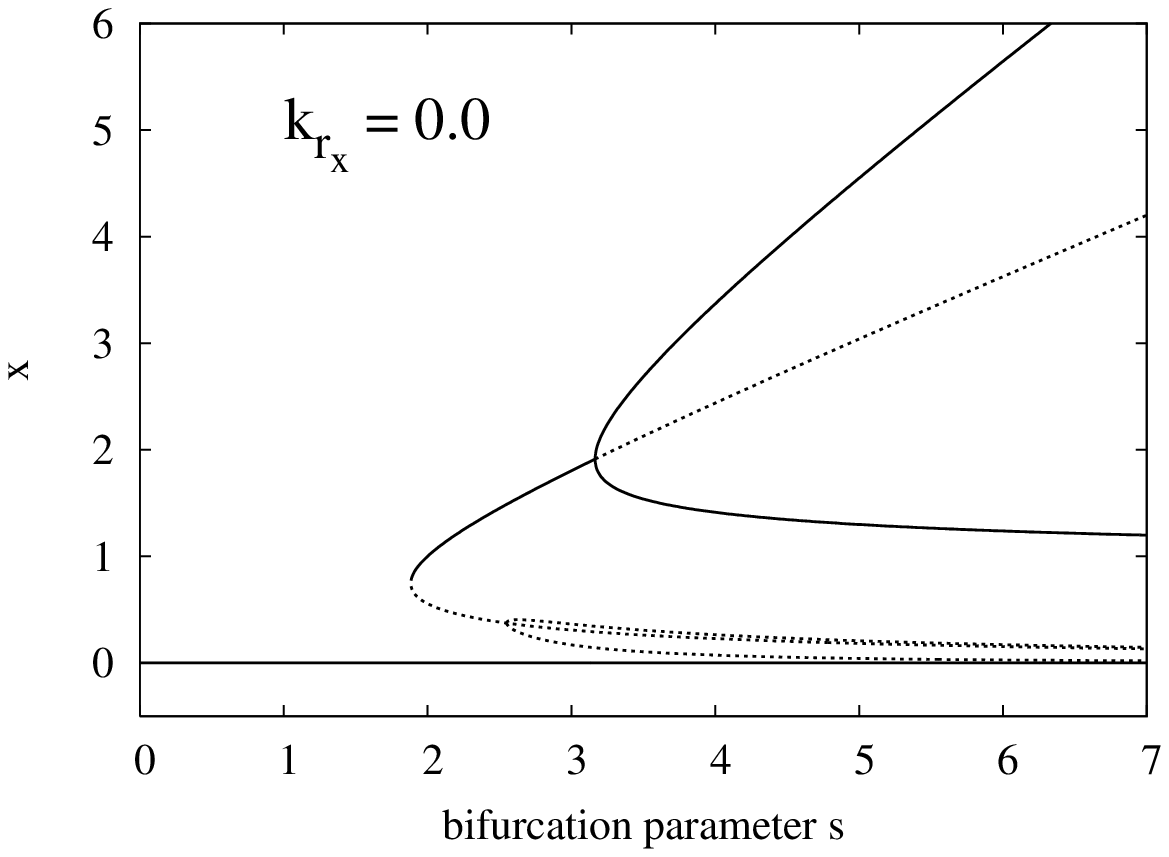}}
\end{minipage}
\begin{minipage}{.32\textwidth}
\subfigure[]{\includegraphics[clip,width=5.0cm]{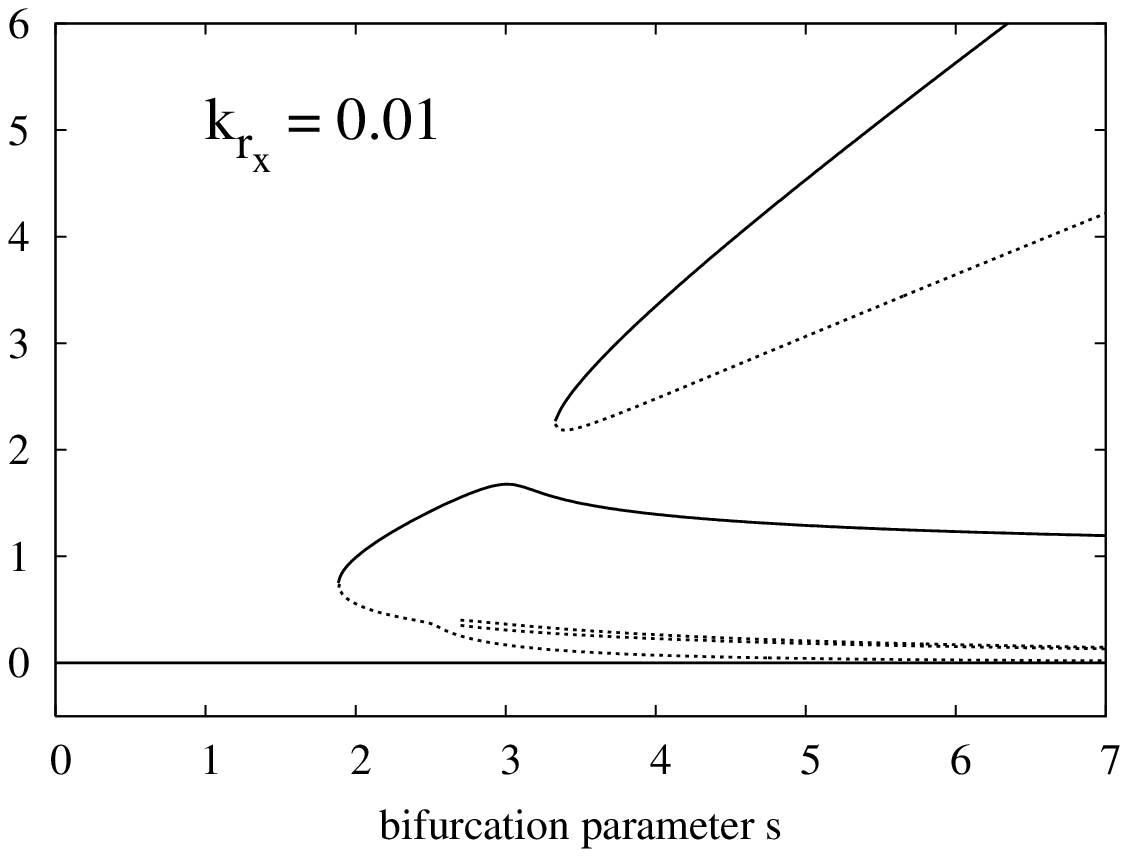}}
\end{minipage}
\begin{minipage}{.33\textwidth}
\subfigure[]{\includegraphics[clip,width=5.0cm]{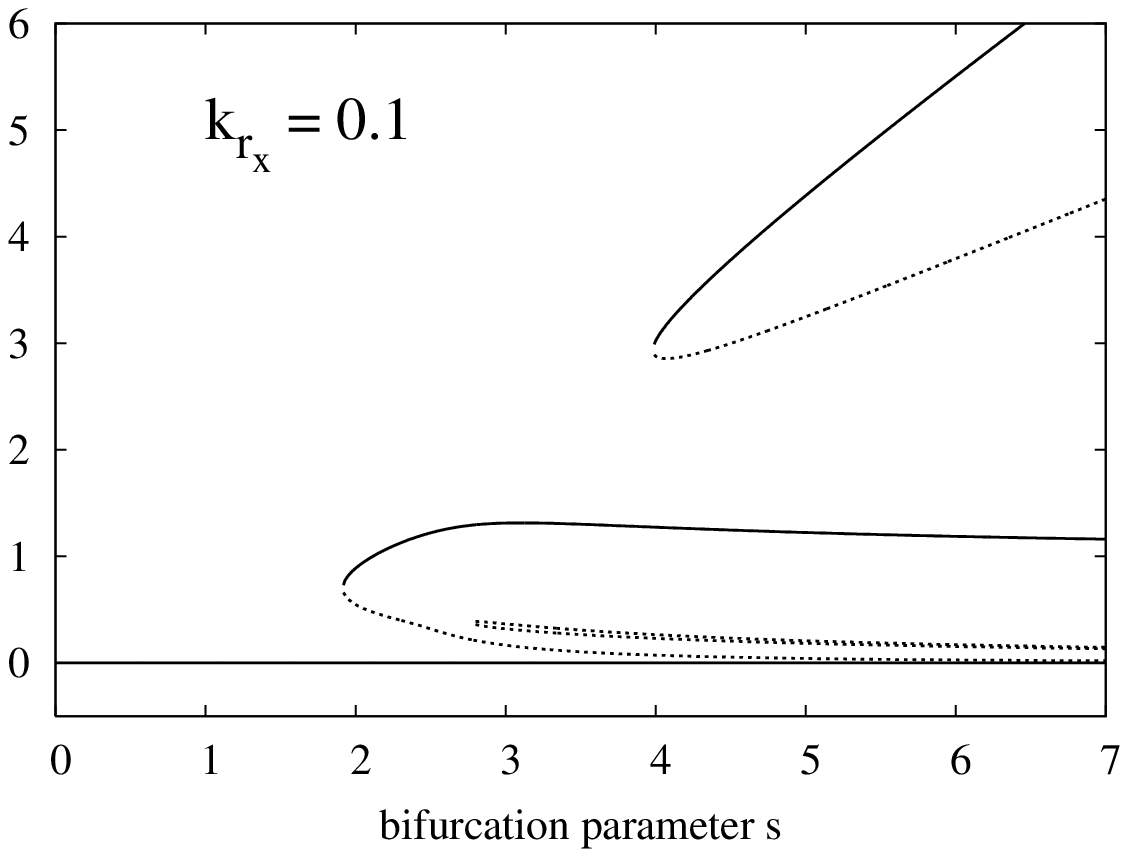}}
\end{minipage}
\end{center}
\caption[Asymmetric scenario]{The deformation of the nullclines for
  the case $k_{r_{x}}=0.1$ is depicted in figures (a) to (c).
  Parameters are $k_{r_{y}}=0$, $u=0.8$, $k_{u}=1$, and $s=1.99$ (a),
  $s=3.0$ (b), $s=6.8$ (c). The trivial fixed point at $(0,0)$ is
  marked by $\blacksquare$, the stable/unstable fixed points off the
  diagonal by $\bullet$/$\circ$. The qualitative change in the
  bifurcation behavior is shown in figures (d) to (f). The bifurcation
  diagrams are shown for increases in the asymmetry parameter
  $k_{r_{x}}$. Parameters are $k_{r_{y}}=0$, $u=1$, $k_{u}=0.8$ and
  $k_{r_{x}} = 0.0$ (d), $k_{r_{x}} = 0.01$ (e), and $k_{r_{x}} = 0.1$
  (f). Solid lines indicate stable, dashed lines unstable fixed
  points. }
\label{fig:ode.bifurcationAsymm}
\end{figure}

For any $k_{r_{x}}>0$ there is a symmetry breaking which shifts the
previously observed bifurcations off the diagonal and destroys the
pitchfork bifurcation observed in the symmetric case for large $u$.
The two up-regulated stable fixed points are not created
instantaneously by the transformation of a previous stable state at
the diagonal, but the initial stable point remains unchanged while a
further (saddle node) bifurcation forms the second up-regulated stable
point alongside with one unstable fixed point. This scenario is shown
in the sequence of nullclines in Fig.
\ref{fig:ode.bifurcationAsymm}(a)-(c). The parameter $k_{r_{x}}$
regulates the distance between the up-regulated stable points and the
extension of their basins of attraction. This is visualized in the
bifurcation diagrams in Fig.  \ref{fig:ode.bifurcationAsymm}(d)-(f)
for different values of $k_{r_{x}}$.

For small unspecific transcription rates $u$, where in the symmetric
case the additional up-regulated stable states are created off the
diagonal, no qualitative changes are introduced by the functional
asymmetry. 

The introduction of asymmetry is not necessarily based on different
interaction mechanisms. It is plausible that auto-regulative
transcription activation does not require identical transcription
rates for the genes of interest. This can be described by relaxing the
symmetry assumption of Section \ref{model:general}, which leads to
gene specific transcription rates $s_x$ and $s_y$.  This asymmetry in
transcriptional activity results in a qualitatively similar symmetry
breaking as in the case of the mechanistic asymmetry: the pitchfork
bifurcation occurring for large $u$ is replaced by a remaining stable
state alongside a saddle-node bifurcation forming the second
up-regulated stable state (data not shown). The magnitude of the
difference in the specific transcriptions rates $s_x$ and $s_y$
regulates the distance between the up-regulated stable states in the
phase plane.

In a scenario where asymmetry of interaction mechanisms occurs
alongside an asymmetry in the specific transcription rates, the
effects on the system behavior combine, either amplifying or
compensating each other.

\subsection{Over-expression scenarios}\label{res:overExpr}

\begin{figure}
\begin{center}
\begin{minipage}{.31\textwidth}
\subfigure[]{\includegraphics[clip,width=5.0cm]{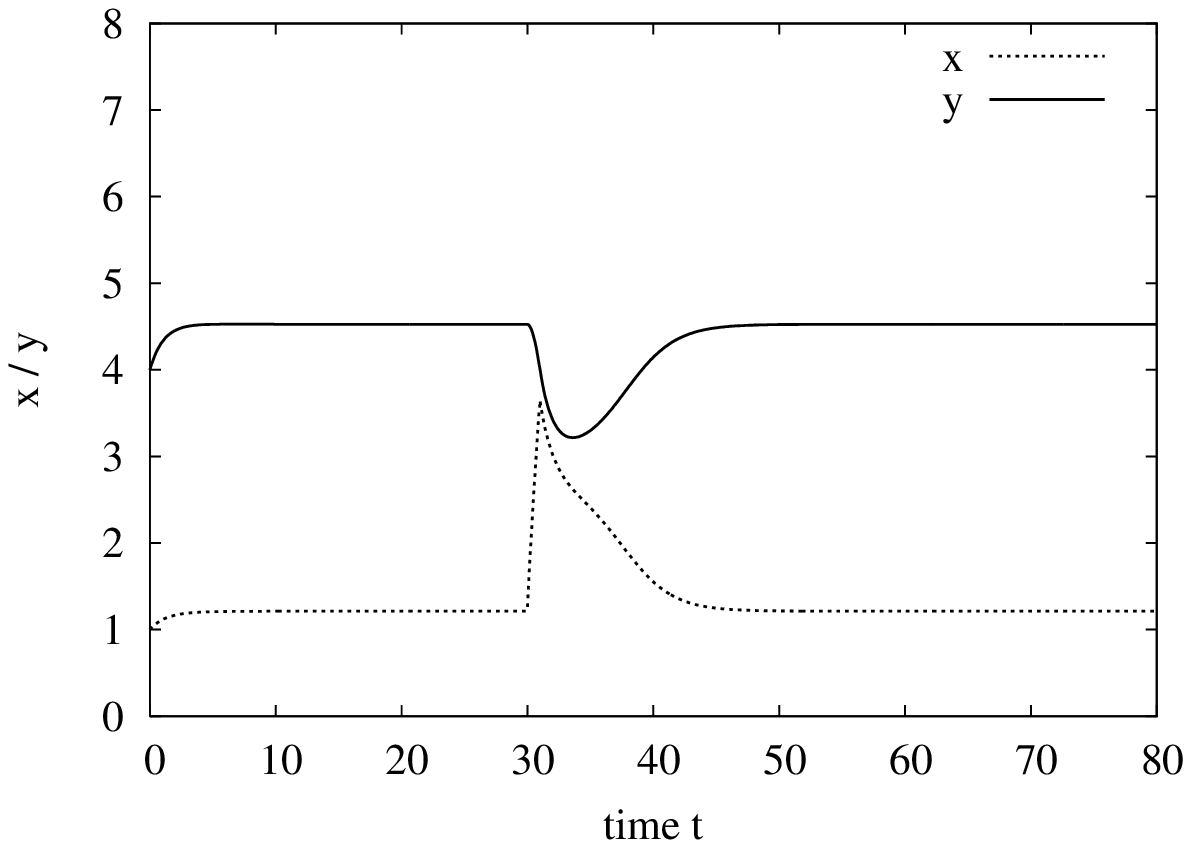}}
\end{minipage}
\begin{minipage}{.31\textwidth}
\subfigure[]{\includegraphics[clip,width=5.0cm]{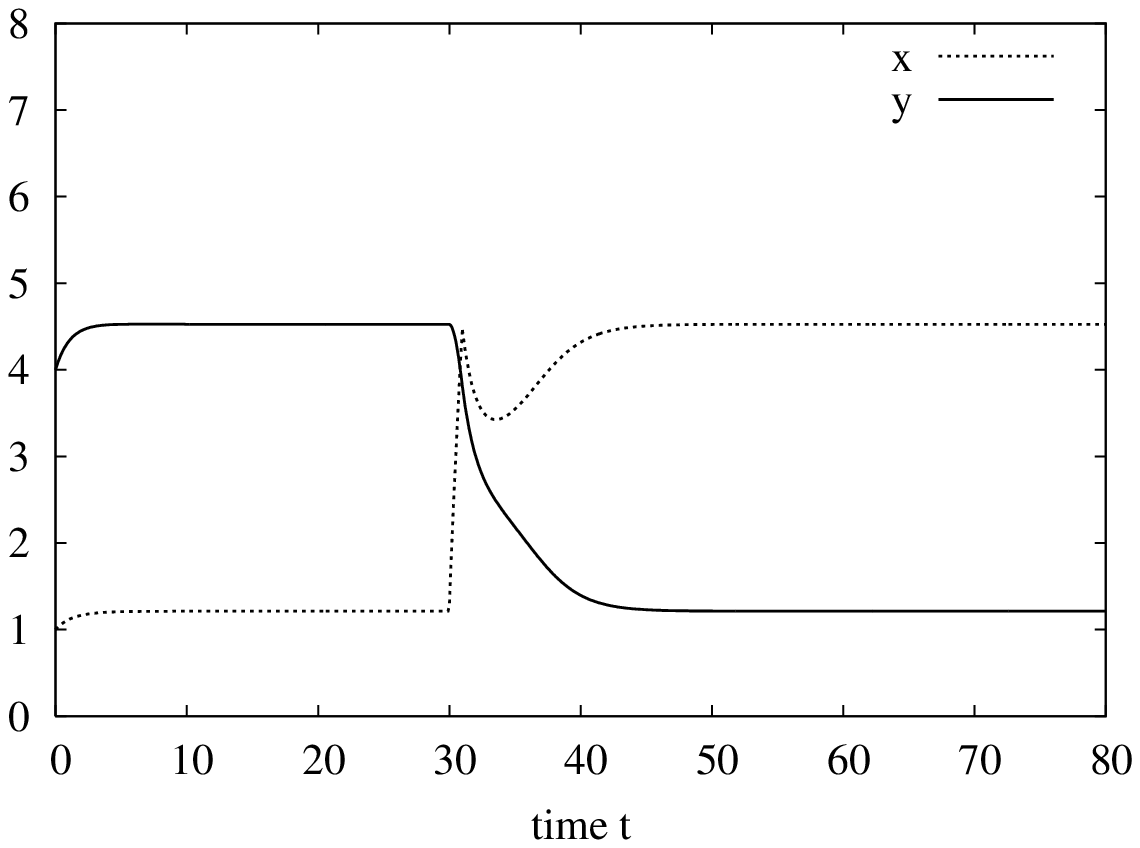}}
\end{minipage}
\begin{minipage}{.31\textwidth}
\subfigure[]{\includegraphics[clip,width=5.0cm]{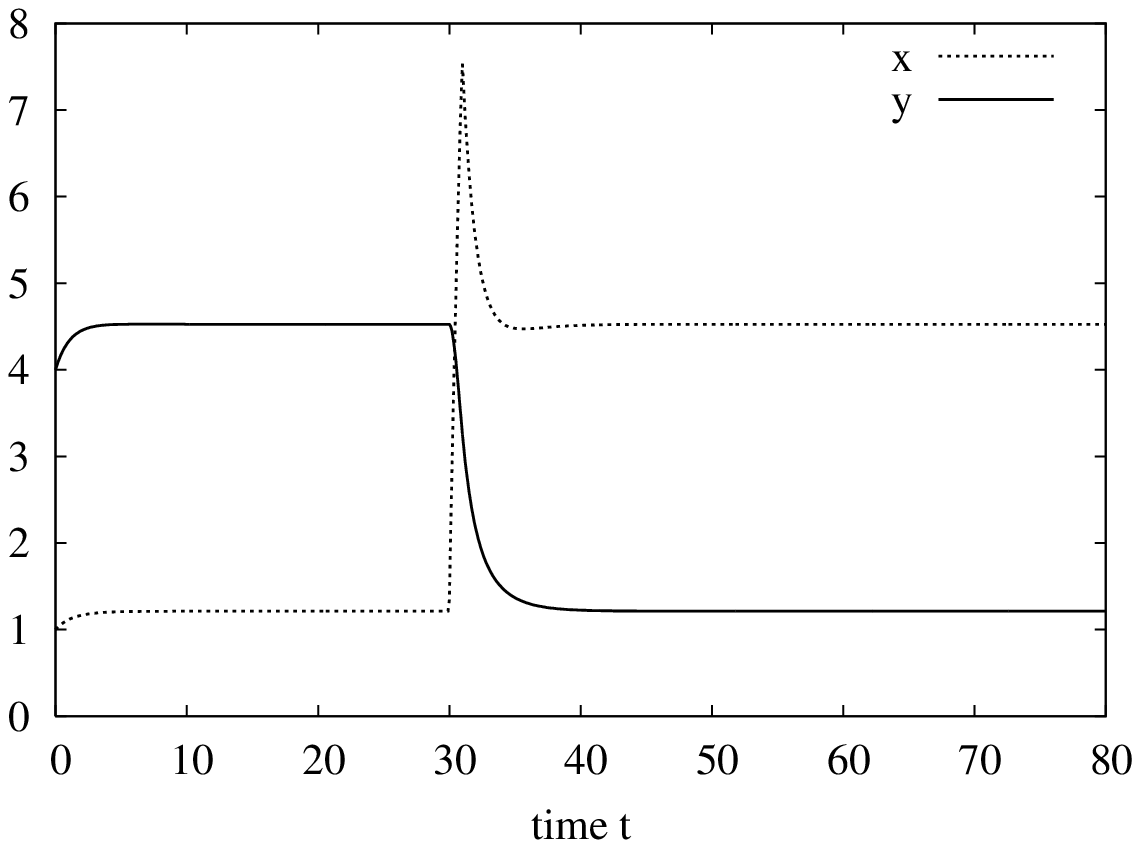}}
\end{minipage}

\begin{minipage}{.31\textwidth}
\subfigure[]{\includegraphics[clip,width=5.0cm]{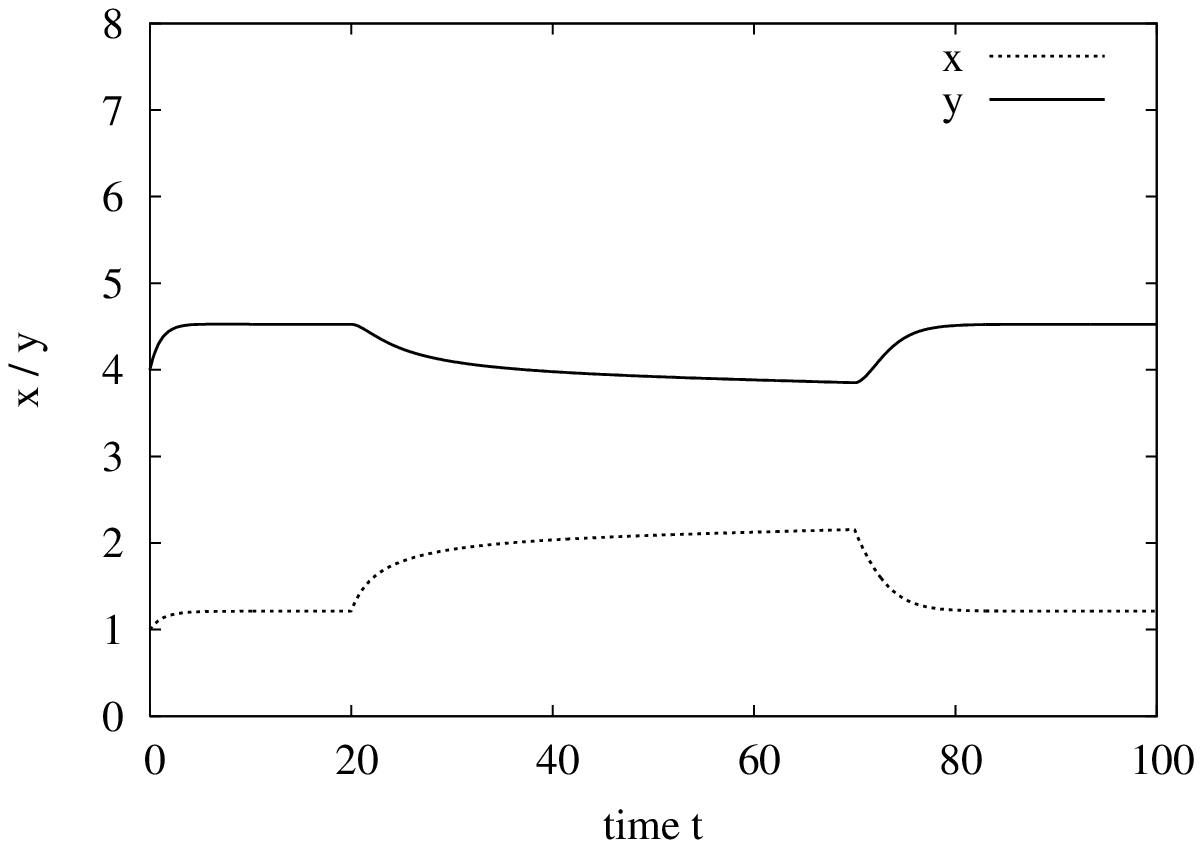}}
\end{minipage}
\begin{minipage}{.31\textwidth}
\subfigure[]{\includegraphics[clip,width=5.0cm]{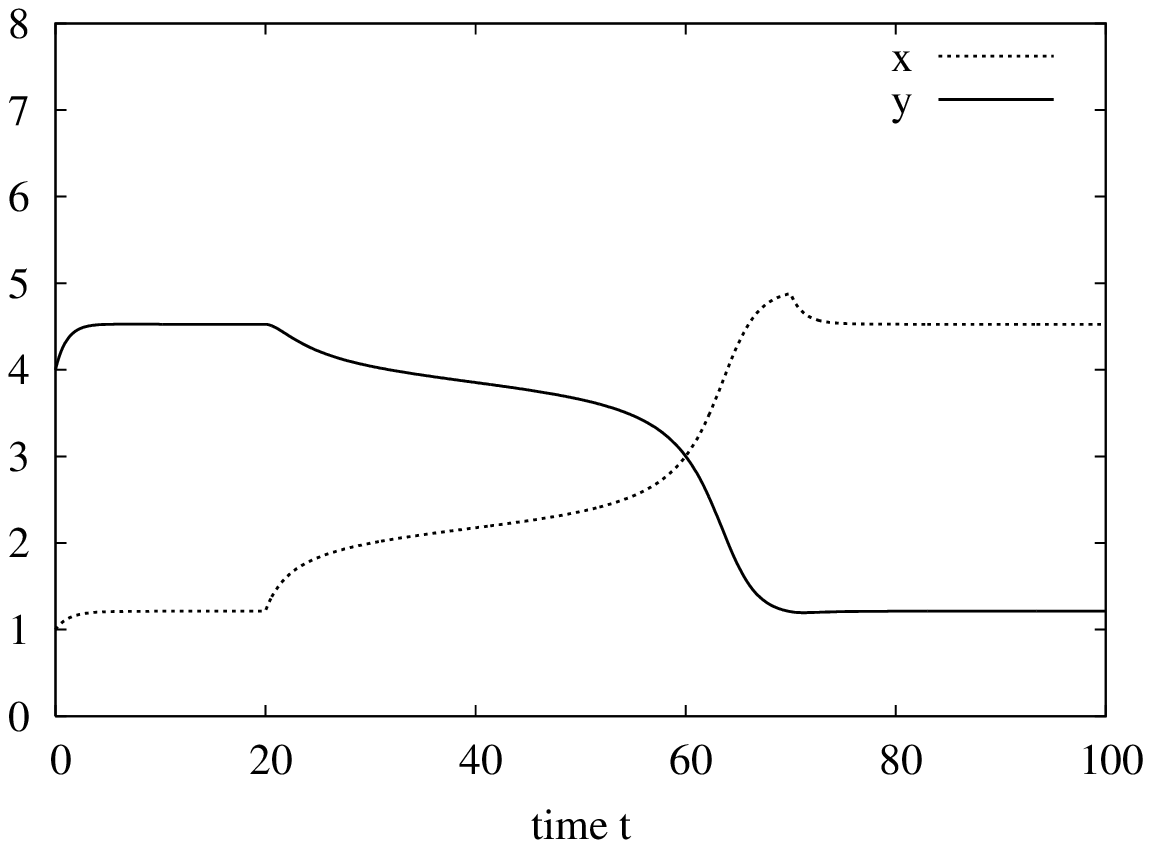}}
\end{minipage}
\begin{minipage}{.31\textwidth}
\subfigure[]{\includegraphics[clip,width=5.0cm]{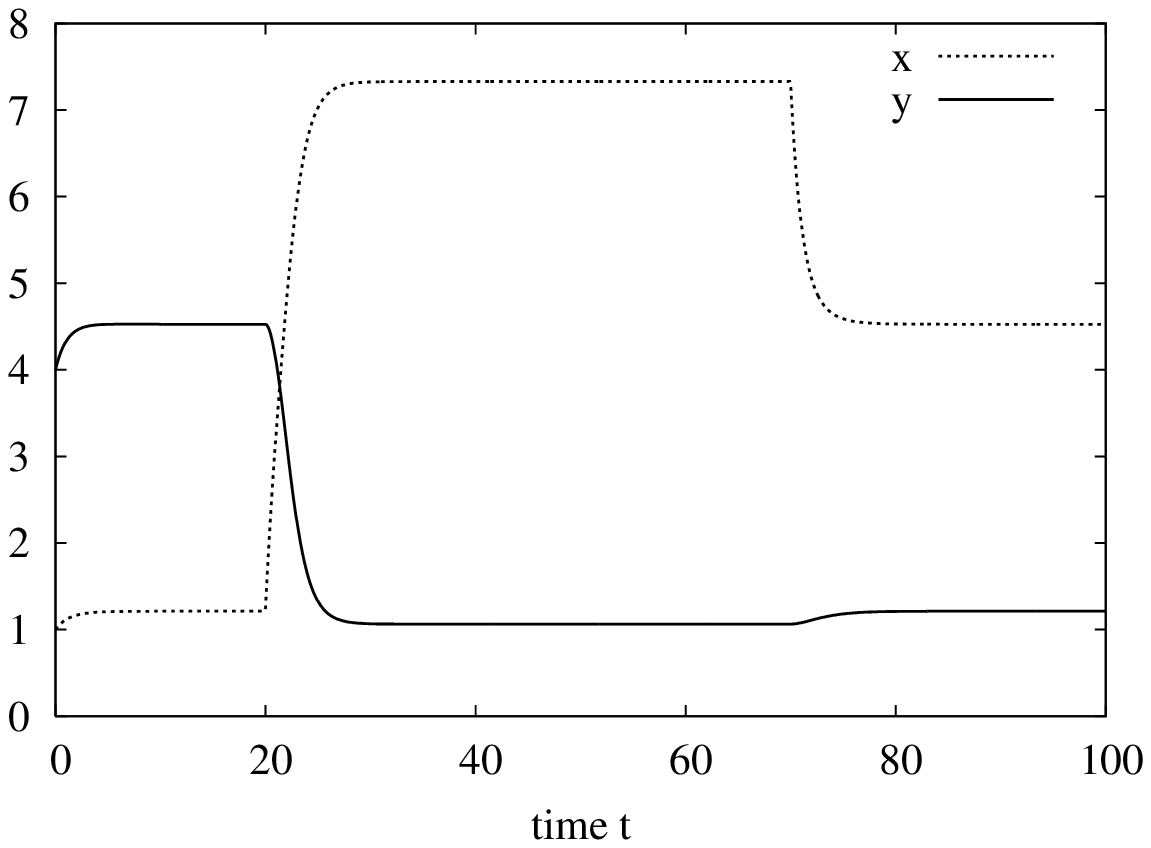}}
\end{minipage}

\caption[Over-expression]{Scenarios for sub- and supercritical
  over-expression. In the subcritical scenarios (a) and (d) the
  transcription factor concentrations remain at the same fixed point,
  whereas in the supercritical scenarios (b),(c),(e) and (f) the
  over-expression leads to a change of the basin of attraction,
  resulting in an different final value of the transcription factor
  concentrations. Over-expression is applied as a short term influence
  at time $t=30$ ((a),(b) and (c)) and long term influence starting at
  time $t=20$ ((d),(e) and (f)).  Parameters are $s=5$, $u=1$,
  $k_{u}=1$, and $k_{r}=0$. The over-expression is applied with
  amplitude $a_{\rm oe}$ and duration $d_{\rm oe}$: $a_{\rm oe}=3$ and
  $d_{\rm oe}=1$ (a), $a_{\rm oe}=4$ and $d_{\rm oe}=1$ (b), $a_{\rm
    oe}=8$ and $d_{\rm oe}=1$ (c), $a_{\rm oe}=0.3$ and $d_{\rm
    oe}=50$ (d), $a_{\rm oe}=0.32$ and $d_{\rm oe}=50$ (e), $a_{\rm
    oe}=2.5$ and $d_{\rm oe}=50$ (f).} \label{fig:overexp}
\end{center}
\end{figure}

Induced over-expression of a certain critical component is a common
experimental method to study interaction dynamics between different
transcription factors and has also been applied to the GATA-1 / PU.1
system \citep{nerlov-c-2000-2543-a, rekhtman-n-1999-1398-a,
  zhang-p-2000-2641-a}. These experiments provide insight in the
stability of the system, interaction time scales, and the role of
co-factors and interaction mechanisms. We have applied an
over-expression impulse of amplitude $a_{\rm oe}$ and duration $d_{\rm
  oe}$ to the model system given in equations \eqref{xeq_sym},
\eqref{yeq_sym}. Characteristics of the dynamic response are only
valid under the outlined steady state assumptions. A qualitative
overview of the simulation results is presented in Fig.
\ref{fig:overexp}.  Starting from a fully symmetric system as studied
in Section \ref{res:symmCase} where, for large $s$, the system is in
one of the two up-regulated states (characterized by one high and one
low expressed transcription factor) two modes of over-expression are
applied: a short impulse over-expression of the lower expressed
component and a long and steady over-expression of the same component.
Not surprisingly, the model reacts to the over-expression with two
distinct scenarios, depending on the intensity of the impulse. For a
subcritical over-expression the system returns to the previous
expression level (indicated in Figs. \ref{fig:overexp}(a) and (d)),
whereas for a supercritical situation the former expression state is
reversed (indicated in Figs \ref{fig:overexp}(b), (c), (e) and (f)).
Translating this picture into the $x$ vs. $y$ phase plane, the
supercritical over-expression corresponds to a change from one basin
of attraction to another, induced by a crossing of the separatrix.
Most available experimental techniques to artificially induce gene
expression lead to a massive over-expression that significantly
exceeds physiological levels, a scenario still underestimated by Figs.
\ref{fig:overexp}(c) and (f). A sensitively tuned expression
experiment is more promising to elucidate critical intensities and
time scales necessary to induce a permanent shift in the genetic
expression patterns and thus to characterize the stability of the
initial states.

\section{Discussion}\label{diss}

The presented model of transcription factor interaction is based on
principles of coupled feedback regulations, which have previously been
proposed for the description of general genetic switches
\citep{becskei-a-2001-2528-a,cinquin-o-2005-233-a,
  francois-p-2004-580-a,gardner-t-2000-339-a, glass-l-1973-103-a} and
the modeling of prokaryotic gene regulation
\citep{mcadams-h-1998-199-a,santillan-m-1998-261-a,
  santillan-m-2001-1364-a,santillan-m-2004-75-a}. Here, specific
experimental knowledge of activation and inhibition mechanisms of two
transcription factors (GATA-1 and PU.1), which play a key role in the
myeloid/erythroid differentiation process of hematopoietic progenitor
cells, is incorporated in this general framework.

Our model analysis particularly focuses on the investigation of the
steady states of transcription factor expression and there dependence
on parameter changes. In this context, we are able to analyze the
experimentally suggested feedback structures and their effects on the
system behavior under various conditions.

To facilitate the mathematical analysis, a number of simplifications
have been made. We interpret the transcription factors described in
the model ($X$ and $Y$) as representatives of a more complex factor
formation rather than an explicit model of PU.1 and GATA-1 alone.
Also, we are aware that most of the statements resulting from the
model analysis are only semi-quantitative in the sense that for all
model parameters, as there are DNA binding-, decay-, and
transcription-rates, no experimentally determined estimates are
available for the investigated system. In the same line of
argumentation, details of the transcription/translation process, like
the DNA binding sequence of transcription factor molecules and the
delay induced by the processes of transcription and translation, have
been excluded from the analysis. Although such phenomena can influence
the dynamics of the system \citep{bundschuh-r-2003-261-a,
  vilar-j-2002-5988-a}, these effects are speculative since detailed
information about relevant rates and time scales are not available.
The simplifications arising from the quasi steady state assumption
outlined in Section \ref{model} for dimerization and DNA binding
impose further limitations on our model with respect to the exact
description of the system dynamics
\citep[c.f.][]{pirone-r-2004-111-a}. However, these simplifications do
not effect the steady state behavior, and, thus, do not alter the
results derived in Section \ref{res}.

The functional role of the so called \emph{priming} behavior is a
question of particular biological relevance which is addressed by this
model. It has been suggested that low level co-expression of multiple
transcription factors, specific for different lineages, might be a
characteristic of (hematopoietic) tissue stem cells
\citep{akashi-k-2005-125-a,cross-m-1997-609-a,orkin-s-2000-57-a}.
However, it is currently unclear whether \emph{priming} corresponds to
a stable state of low level co-expression or to a truly
zero-expression overlaid by some random expression noise. Furthermore,
there is a hypothesis that lineage specification induction might be a
two stage process with a primary initialization of transcription
factor network interaction (i.e., a transition from no expression to
low level co-expression) and a secondary network-induced
differentiation process \citep{enver-t-1998-9-a}.
This perspective immediately leads to the questions under which
conditions such a two stage process can be established and whether
such a sequence of different activation states of the transcription
factor network requires (multiple) external induction signals or
whether it represents a system inherent development.

The suggested model generates two characteristic modes of system
stability depending on the magnitude of the specific transcription
rate $s$: For small $s$ only the trivial fixed point $(0,0)$ exists;
for large $s$ two additional up-regulated stable states are observed
that are marked by the dominance of one factor over the other
(dominated co-expression).  These modes are maintained independently
of a mechanistic or parametric asymmetry. Assuming a differentiation
initiation by increasing the transcription rate $s$ (e.g. by changes
in chromatin structure \citep{berger-s-2001-263-a,
  rosmarin-a-2005-131-a} or by alterations in activation/inhibition
complexes \citep{hume-d-2000-2323-a}), the transition between the
different stable states is the central mechanism characterizing
lineage specification.

Within the proposed biological framework the trivial fixed point at
$(0,0)$, which exists for all values of $s$, can be identified with
the undifferentiated state of a cell where neither activation nor
decision processes are observed.  It should be mentioned that
stability of this fixed point is specific for the outlined model and
has not been observed for the general case of a toggle switch
\citep[c.f.][]{gardner-t-2000-339-a}. In logical extension, the two
up-regulated stable fixed points, observed for large $s$, would be
interpreted as expression states promoting one or the other lineage.
These distinct states are characterized by a high auto-regulative
expression of one dominating factor and a reduced expression of the
antagonistic factor.

In Sections \ref{res:symmCase} and \ref{res:asymm} it has been
demonstrated that, for increased unspecific transcription rates $u$
(but only in this case!), a further stable fixed point exists for
intermediate $s$ prior to the formation of the two distinct states of
dominated co-expression. This particular fixed point is characterized
by a balanced low level co-expression of the two antagonistic factors
where no final commitment decision has been made. The resulting
transition sequence between three distinct regions of multi-stability
can be interpreted as a possible explanation for a two stage
differentiation process mentioned above.
 
The induction of a system change from the stable trivial fixed point
to the dominated or, if existent, balanced low level co-expression
state, needs to be triggered either by a stochastic background
expression or by an active impulse on the system. The unstable fixed
point separating the trivial from the up-regulated stable states is an
indicator of the size of the basins of attraction. The observation
that the unstable fixed point approaches the trivial one for
increasing $s$ indicates that the magnitude of the perturbation to
introduce a transition from the zero-state to the co-expression states
decreases in the same fashion: for a sufficiently large $s$ even a
small perturbation is able to initiate differentiation.

Concluding from these results, there are two different scenarios to
explain the experimentally suggested \emph{priming} behavior within
the proposed model framework: (1) \emph{Priming} might be considered
as the existence of perturbations in the expression of transcription
factors, imposed on a zero-expression state represented by the trivial
fixed point at $(0,0)$, either in the form of stochastic background
fluctuations (functional noise) or by active impulses. In this
scenario, the perturbations are necessary components of the regulatory
system to induce a differentiation process. It points to the potential
role of stochastic effects in the context of decision making in stem
cell differentiation as frequently suggested (see
\citet{kaern-m-2005-451-a} for a review). (2) In contrast to this
scenario, \emph{priming} can also be explained by the balanced low
level co-expression state, which becomes unstable for increasing
specific transcription rates. Due to this parameter dependent loss of
stability, this scenario would lead to differentiation without the
need for external perturbations\footnote{To be precise: An
  infinitesimal perturbation is required to escape from the unstable
  fixed point. Fluctuations of this magnitude are present in any
  ``real world'' system.}. However, the balanced low level
co-expression state is only existent if there is a certain degree of
unspecific transcription.

Currently, our results do not allow to decide between the two
scenarios. The introduction of artificial differentiation impulses of
different intensities on uncommitted cells might be an appropriate way
to tackle this question experimentally. Whereas, a low level
co-expression \emph{priming} (like in scenario (2)) would be unaffected
by these perturbations, the system could be enforced to escape the
\emph{priming} status in scenario (1). Moreover the existence and the
stability of the different stable system states depend sensitively on
the model parameters. Due to the lack of available data on
transcription and binding rates, we are currently not able to specify
the biological relevant regimes more rigorously.  Any experimental
approximation of binding and transcription rates for the involved
components supports the identification of the nature of
\emph{priming}.

The over-expression scenarios presented in Section \ref{res:overExpr}
fail to explain experimental findings described by several authors
\citep{nerlov-c-2000-2543-a,rekhtman-n-2003-7460-a,zhang-p-2000-2641-a}.
In spite of the induced up-regulation of one transcription factor it
was observed that the transcription level of the antagonistic
transcription factor remained more or less constant.  These
observations are in contrast to the model results presented here, in
which the induced over-expression of the initially low expressed
factor shifts the equilibrium to the opposing co-expression state.
Retaining our model assumptions, a potential interpretation can be
given as follows: One of the major functions of transcriptional
regulators like GATA-1 and PU.1 is the activation of a set of
lineage-specific genes which include further transcription and growth
factors as well as functional components of the committed lineages
\citep{tenen-d-2003-89-a}. In \citet{sieweke-m-1998-545-a} and
\citet{tsai-s-1991-919-a}, the authors point to a continuously
modulated set of cooperative lineage-inherent transcription factors
changing with the state of differentiation.  Such secondary complexes
of transcription factors could in turn act as activators of the
initial transcription factor, substituting for a simple
auto-regulation and thus stabilizing the initial up-regulation pattern
\citep{hume-d-2000-2323-a}.  In such a scenario our model would only
account for the initial switching process.  The experimentally
observed stable transcription level of the antagonistic factor in
over-expression experiments could be interpreted as a substitution of
the auto-regulation by secondary transcription factor complexes.

Summarizing, the presented model is able to provide a quantitative
explanation for possible mechanisms underlying lineage specification
control in eukaryotic systems. It is able to generate parameter
dependent changes in the system behavior, with alteration of the
number of possible stable steady states. Specifically, the model
explains states of stable co-expression as well as the situation
characterized by an over-expression of one factor over the other. The
conditions inducing shifts from one to another stable state (e.g.
parameter choice, degree of system disturbances), however, depend in a
sensitive manner on the assumed activation and inhibition mechanisms.
Using the mathematical model, we were able to test several
combinations of experimentally described feedback mechanisms with
respect to their influence on the resulting stable states and provide
possible explanations for the experimentally suggested differentiation
\emph{priming} of stem cells.

\section*{Acknowledgment}

The authors thank Michael Mackey for his encouragement and critical
discussions in the process of preparing this manuscript, and Michael
Cross for his explanations of many biological details. Furthermore we
acknowledge the Human Frontier Science Program which supported
initiation of this work by short-term fellowship to I.R. at the Centre
for Nonlinear Dynamics, McGill University, Montreal.

\appendix
\section{Derivation of transcription factor dynamics}\label{app:TF_dynamics}

It is assumed that the transcription of transcription factors $X$
and $Y$ requires the existence of activator complexes, i.e., the
binding of $X$ or $Y$ dimers to the promoter regions of $X$
($D_x$) and $Y$ ($D_y$), respectively. As described in Section
\ref{model:ode}, we distinguish between a specific (see
equations \eqref{xsprod},\eqref{ysprod}) and an unspecific
(equations \eqref{yuprod},\eqref{xuprod}) transcription activation.
Furthermore, there is the possibility that $X$ and $Y$ can act
jointly as a repressor dimer $Z_1$, inhibiting the DNA binding of the
$X$ and $Y$ activator dimers (see equations
\eqref{zrepx},\eqref{zrepy}).

The total amount of promoter sites for $X$ and $Y$ can be
specified as the sum of unbound (free) and occupied (by repressor
or activator molecules) promoter regions, i.e.,
\begin{equation}
D_{x/y}^{tot}=D_{x/y}+D_{x/y}^{xy}+D_{x/y}^{xx}+D_{x/y}^{yy} \quad .
\end{equation}
Using the equilibrium (dissociation) constants
\begin{multline}
K_1=\frac{D_x^{xx}}{D_x x^2}, \quad K_2=\frac{D_x^{yy}}{D_x y^2}, 
\quad K_3=\frac{D_y^{yy}}{D_y y^2}, \quad K_4=\frac{D_y^{xx}}{D_y x^2},\\
K_6=\frac{D_x^{xy}}{D_x xy},\mbox{ and} \quad K_7=\frac{D_y^{xy}}{D_y xy} \, ,
\end{multline}
obtained from assuming equations \eqref{xsprod}-\eqref{xuprod},
\eqref{zrepx}, \eqref{zrepy} to be in a quasi steady state, the
fraction of promoter sites contributing to active $X$ and $Y$
transcription is given by
\begin{equation}
\begin{split}
\frac{D_x^{xx}+D_x^{yy}}{D_x^{tot}} 
& =\frac{K_1 x^2 D_x + K_2 y^2 D_x}{D_x + K_1 x^2 D_x + K_2 y^2 D_x + K_6 xy D_x}\\
& =\frac{K_1 x^2 + K_2 y^2}{1 + K_1 x^2 + K_2 y^2 + K_6 xy}
\end{split}
\end{equation}
and
\begin{equation}
\begin{split}
\frac{D_y^{yy}+D_y^{xx}}{D_y^{tot}}
& =\frac{K_3 y^2 D_y + K_4 x^2 D_y}{D_y + K_3 y^2 D_y + K_4 x^2 D_y + K_7 xy D_y}\\
& =\frac{K_3 y^2 +K_4 x^2}{1 + K_3 y^2 + K_4 x^2 + K_7 xy}
\end{split}
\end{equation}
respectively.

Taking the (first order) decay rates of $X$ and $Y$ into account,
one immediately obtains equations \eqref{xeq_compl},
\eqref{yeq_compl} by writing down the balance equations for $X$
and $Y$.

\section{Domain of the nullclines}\label{app:deformation}

Under the equilibrium assumption equation
\eqref{xeq_sym_equilib} can be solved for:
\be
\yy_{1/2}(\xx) = \frac{k_{r} \xx^2 \pm \sqrt{k_{r}^2 \xx^4 - 4 k_{u} \xx (u-\xx) (s \xx-1 - \xx^2)}}{2 k_{u} (u-\xx) }
\ee
which describes a set of nullclines. There is an obvious
singularity at $\xx=u$.

\begin{figure}
\begin{center}
\begin{minipage}{.48\textwidth}
\subfigure[]{\includegraphics[clip,width=7cm]{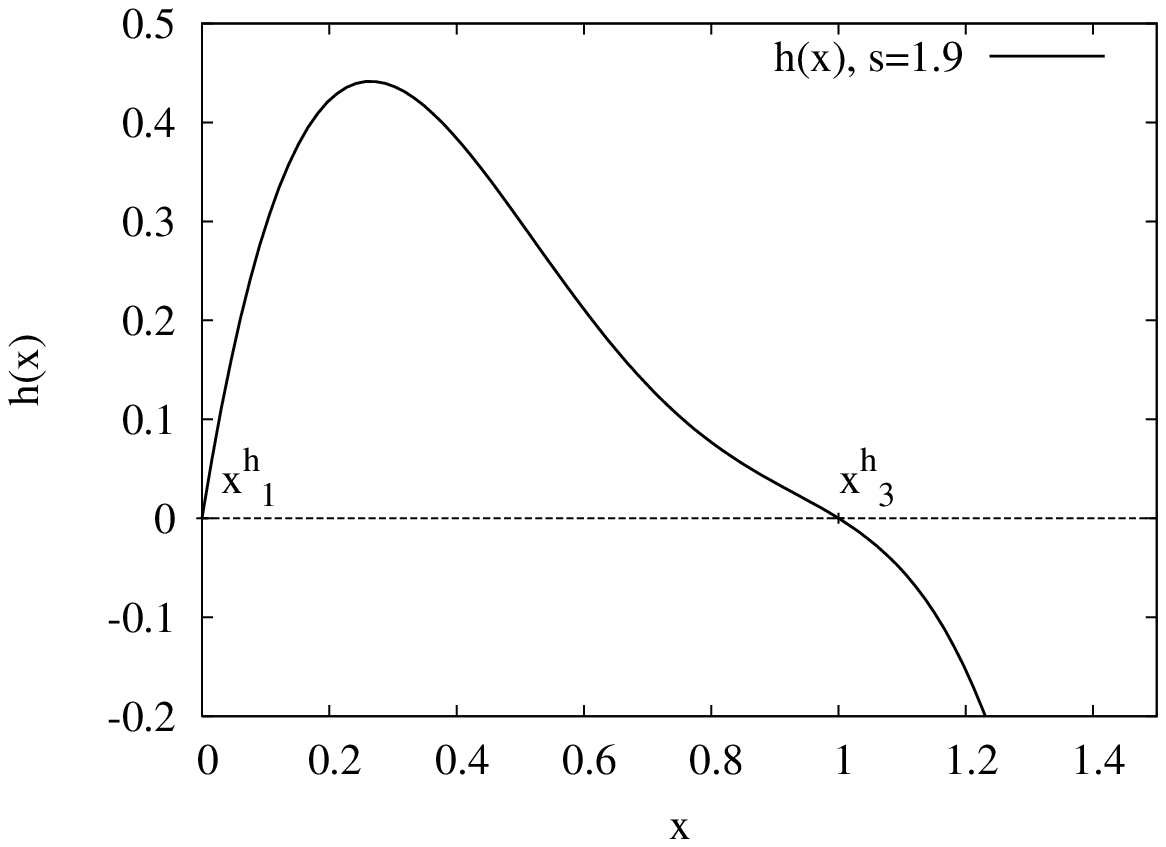}}
\end{minipage}
\begin{minipage}{.47\textwidth}
\subfigure[]{\includegraphics[clip,width=7cm]{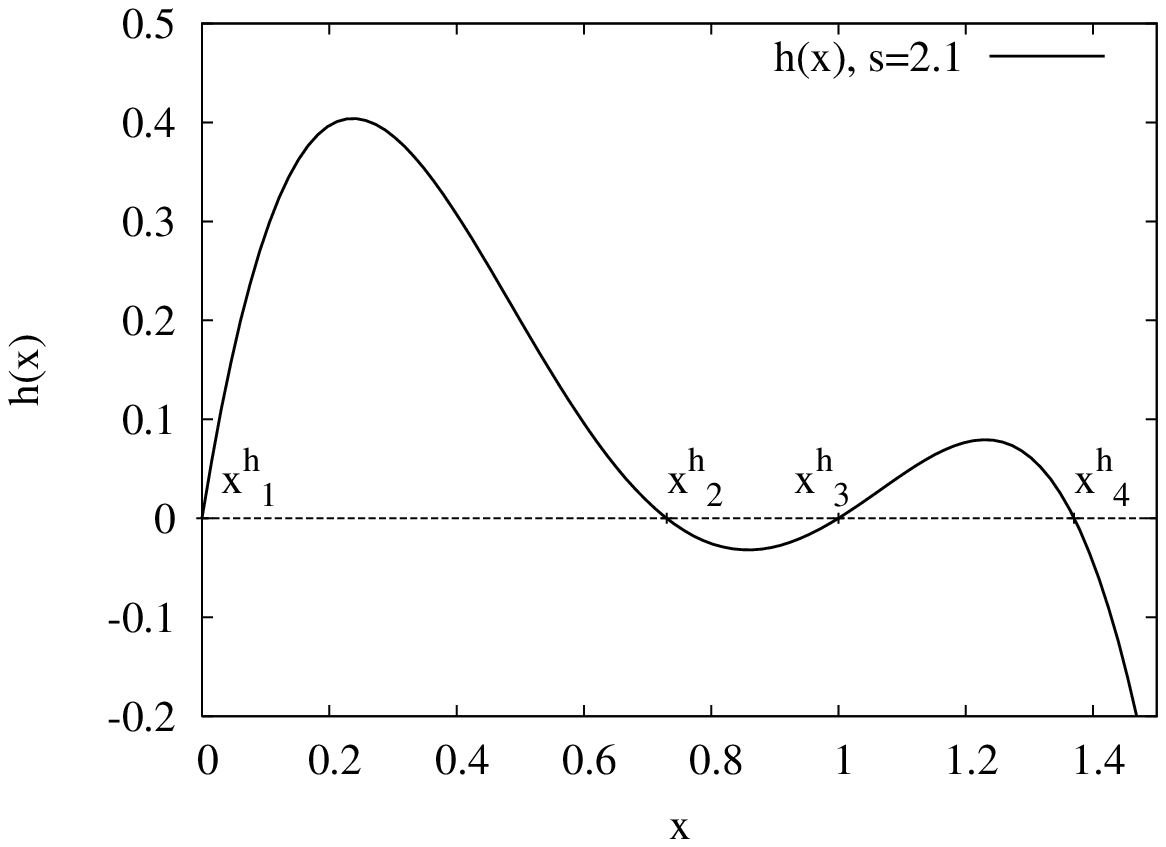}}
\end{minipage}
\caption[Domain of the nullclines]{The function
$h(\xx)= k_{r}^2 \xx^4 - 4 k_{u} \xx (u-\xx) (s \xx-1 -\xx^2)$ is
shown for $k_{r}=0$, $u=1$, $k_{u}=1$ and $s=1.9$ (a) and $s=2.1$
(b).} \label{fig:defSpaceNC}
\end{center}
\end{figure}

The solutions $\yy_{1/2}(\xx)$ are real for $0 < h(\xx) = k_{r}^2
\xx^4 - 4 k_{u} \xx (u-\xx) (s \xx-1 -\xx^2)$ with $h(\xx)$ defined as
the expression under the square root in the previous equation. For
$k_{r}=0$ the roots of $h(\xx)$ are located at
\begin{align}
\xx^{h}_{1} &= 0\\
\xx^{h}_{2} &= \frac{s}{2}-\sqrt{\frac{s^{2}}{4}-1}\\
\xx^{h}_{3} &= u\\
\xx^{h}_{4} &= \frac{s}{2}+\sqrt{\frac{s^{2}}{4}-1}
\end{align}
Real roots at $\xx^{h}_{2/4}$ exist only for $s\geq2$. Fig.
\ref{fig:defSpaceNC} shows the function $h(\xx)$ for $s=1.9$ and
$s=2.1$. In the case $s<2$ the parameter $u=\xx^{h}_{3}$ restricts the
definition space of the nullclines to $\xx \in [0,\xx^{h}_{3}]$. For
$s>2$ three scenarios exist, where the singularity at $\xx = u =
\xx^{h}_{3}$ marks a boundary for distinct intervals in the domain:
$\xx^{h}_{3} < \xx^{h}_{2}$ (with $\xx \in
[0,\xx^{h}_{3}] \cup [\xx^{h}_{2},\xx^{h}_{4}]$), $\xx^{h}_{2} <
\xx^{h}_{3} < \xx^{h}_{4}$ (with $\xx \in
[0,\xx^{h}_{2}] \cup [\xx^{h}_{3},\xx^{h}_{4}]$) shown in Fig.
\ref{fig:defSpaceNC}(b), and $\xx^{h}_{4} < \xx^{h}_{3}$ (with $\xx \in
[0,\xx^{h}_{2}] \cup [\xx^{h}_{4},\xx^{h}_{3}]$).

\section{Derivation of bifurcation condition}\label{app:bifurcation}

The nullclines of the symmetric system derived in \eqref{xeq_sym_equilib} and \eqref{yeq_sym_equilib} are interpreted as functions of $\xx$ and $\yy$:
\begin{align}
\xx &= f(\xx,\yy) = \frac{s \xx^2 + u k_{u} \yy^2}{1+ \xx^{2} + k_{u} \yy^{2}+ k_{r} \xx\yy}, \label{xeq_sym_equilibAPP}\\
\yy &= g(\xx,\yy) = \frac{s \yy^2 + u k_{u} \xx^2}{1+ k_{u} \xx^{2} + \yy^{2}+ k_{r} \xx\yy}. \label{yeq_sym_equilibAPP} 
\end{align}

To derive bifurcation conditions one has to determine the point of
tangency of the nullclines $f(x,y)$, $g(x,y)$ at a steady state $(x^*,
y^*)$, i.e.
\begin{equation}
\left. \frac{df(x,y)}{dy} \right|_{(x^*,y^*)}=\left. \frac{dg(x,y)}{dx} \right|_{(x^*, y^*)}.
\end{equation}

Generally, it holds for inverse functions $h$ and $k=h^{-1}$ that
$k^{\prime}(h(x))=\left({h^{\prime}(x)}\right)^{-1}$. Considering
only points at the diagonal $x=h(x)=y$, it follows that
$h^{\prime}(x)=\left(k^{\prime}(x)\right)^{-1}$. Assuming identity
of the first order derivatives $h^{\prime}$ and $k^{\prime}$ at
some point $x^*$ on the diagonal yields, therefore,
$\left(h^{\prime}(x^*)\right)^2=\left(k^{\prime}(x^*)\right)^2=1$.

From these statements, it follows that we have to consider the
following equalities to find the bifurcation conditions for the
symmetric system, restricting to symmetric steady states of the
form $(x^*,x^*)$:
\begin{equation}
\left. \frac{df(x,y)}{dx} \right|_{(x^*,x^*)}=\left. \frac{dg(x,y)}{dx} \right|_{(x^*,x^*)}=|1|.
\end{equation}

\bibliographystyle{elsart-harv.bst}

\end{document}